\documentclass[journal,twoside,web]{ieeecolor}
\usepackage{generic_nojournal}
\usepackage{xr}
\usepackage{cite}
\usepackage{amsmath,amssymb,amsfonts}
\usepackage{algorithmic}
\usepackage{graphicx}
\usepackage{textcomp}
\usepackage{cuted,empheq,ulem}
\usepackage{url}
\usepackage{color}

\newcommand{\pdpd}[2]{\frac{\partial #1}{\partial #2}}
\renewcommand{\Re}{\operatorname{Re}}
\renewcommand{\Im}{\operatorname{Im}}

\newcommand{\CC}{{\mathbb{C}}}

\newcommand{\ignore}[1]{}
\usepackage{fancyhdr}
\usepackage{pdfpages}
\DeclareMathOperator*{\argmax}{arg\,max}
\DeclareMathOperator*{\argmin}{arg\,min}
\def\BibTeX{{\rm B\kern-.05em{\sc i\kern-.025em b}\kern-.08em
    T\kern-.1667em\lower.7ex\hbox{E}\kern-.125emX}}
    
\begin{document}

\title{Understanding the combined effect of $k$-space undersampling
  and transient states excitation in MR Fingerprinting reconstructions}
\author{Christiaan C.\ Stolk and Alessandro Sbrizzi
\thanks{Manuscript submitted on October 5, 2018.}
\thanks{Christiaan C.\ Stolk is with the University of Amsterdam, Science Park 107, 1098 XG, Amsterdam, The Netherlands (e-mail: C.C.Stolk@uva.nl).}
\thanks{Alessandro Sbrizzi is with the University Medical Center, Utrecht, Heidelberglaan 100, 3584 CX, Utrecht, The Netherlands (e-mail: a.sbrizzi@umcutrecht.nl).}
}

\maketitle

\begin{abstract}
Magnetic resonance fingerprinting (MRF) is able to estimate multiple quantitative  tissue parameters from a relatively short acquisition. 
The main characteristic of an MRF sequence is the simultaneous
application of (a)~transient states excitation and (b)~highly
undersampled $k$-space.  Despite the  promising empirical results
obtained with MRF, no work has appeared that formally describes the
combined impact of these two aspects on the reconstruction accuracy.
In this paper, a mathematical model is derived that
directly relates the time varying RF excitation and the $k$-space sampling 
to the spatially dependent reconstruction errors.
A subsequent in-depth
analysis identifies the mechanisms by which MRF sequence
properties affect accuracy, providing a formal explanation of several
empirically observed or intuitively understood facts.
New insights are obtained which show how this analytical framework
could be used to improve the MRF protocol.
\end{abstract}


\section{Introduction}
Magnetic resonance fingerprinting (MRF)
\cite{MaEtAl2013_MagneticResonanceFingerprinting,JiangEtAl2015_MRFingerprinting,cloos2016multiparametric},
aims at quantitatively reconstructing multiple tissue parameters from
a relatively short sequence during which the magnetization is in the
transient states. Imaging is performed between two excitation pulses
by means of a strongly under-sampled read-out scheme, for instance
single-shot spirals or few radial spokes. As a consequence, the
reconstructed snapshots exhibit strong Fourier aliasing artifacts
which can be filtered-out by a subsequent matching filter step to
recover the parameters of interest, typically the relaxation times
($T_1$, $T_2$) and the proton density ($\rho$). While different
approaches have been proposed for the reconstruction of MRF data
\cite{DaviesEtAl2014_CS_MRF,McGivneyEtAl2014_SVDCompression_MRFingerprinting,doneva2017matrix,asslander2018low,zhao2016maximum,sbrizzi2017dictionary},
they all rely on (I) transient state acquisitions, (II) some form of
non-uniform Fourier transformation to the spatial domain
\cite{greengard2004accelerating,fessler2003nonuniform} and, with the
exclusion of  \cite{sbrizzi2017dictionary}, (III) look-up table match.

While the empirical results of MRF are promising (as witnessed by the
popularity of the method
\cite{cauley2015fast,chen2016mr,hamilton2017mr}) there is, at the time
of writing, not much work dedicated to the analysis of the error in
the reconstructions. In particular, there is no theory providing
quantitative information on the errors
 originating from the interplay of transient states excitation
and $k$-space undersampling. To tackle the
difficulty of the problem, over-simplistic assumptions are usually
made. For example, the Fourier aliasing artifacts are usually
considered to be (a) independent on the parameter maps, (b)
identically and independently distributed (i.i.d.) and (c) having zero
mean. Some sensitivity studies have been performed on the basis of
this assumption and standard statistical techniques such as the
Cramer-Rao lower bound have been applied
\cite{zhao2017optimal,asslander2017relaxation,sbrizzi2017dictionary}. However,
these simplifications are not a satisfactory choice for an in-depth
analysis and understanding of MRF because: (1) undersampling artifacts
are correlated with the parameter maps; (2) images reveal structure,
which thus is reflected in the nature of the Fourier aliasing meaning
that the underlying parameter maps are also structured; (3) the
intensity of undersampling perturbations depends on the specific
moment during the transient sequence or, in other words, the noise in
the images is heteroscedastic; (4) the size of the Fourier aliasing artifacts and the extent to which they can be assumed to be zero-mean, i.i.d.\ and normally distributed depend on the experimental settings, hence for the sake of generalization it is better not to make these restrictive assumptions.

Clearly, a deeper understanding of the MRF error in relation to experimental design choices is urgent and important; this is the aim of our work. Leveraging on techniques from perturbation theory, we derive a mathematical model which explicitly relates the {\it combined} effect of RF excitation train and $k$-space under-sampling scheme to the systematic error in the reconstructed parameter maps. No assumptions are made with regard to the distribution of the noise terms which are instead treated for what they are, that is, Fourier aliasing perturbations. 

Based on our model, we are able to identify the situations when MRF
works, in the sense that the errors in the reconstructed parameters
are sufficiently small to be reasonably ignored. We also describe the
scenarios for which MRF fails and provide mathematical justifications
for that. In addition, we address the role of randomness and structure
in the $k$-space acquisition and RF excitation train obtaining results
which are somehow unexpected from an intuitive point of view. We also
indicate how the analytical techniques introduced in this work could
be leveraged to further improve the MRF protocol.

The paper is organized as follows. Section \ref{sec:err_model} introduces the basic concepts
and describes the MRF signal model. Section \ref{sec:analysis1}
proceeds with the perturbation theoretic analysis of the error and
identifies the terms which are responsible for the dominant
biases. This is the main innovation of our work and culminates with
equations (\ref{eq:theta1_with_errors}) and (\ref{eq:theta1_with_errors_varrho0}). In Section \ref{sec:tests}, several tests  are reported to validate the proposed model in realistic acquisition/reconstruction scenarios. Subsequently, in Section \ref{sec:in-depth_analysis} we derive general insights into various aspects of MRF such as the role of randomness, cross-talk effects between parameters and transient RF excitation. Finally, a general discussion is provided in Section \ref{sec:discussion}. 

\section{The MRF framework}\label{sec:err_model}
\subsection{Signal model for under-sampled $k$-space acquisitions}
In this section we introduce a model for the MRF  signal in the case of two-dimensional imaging.  The acquisition sequence contains $N_{\rm I}$ radiofrequency (RF) pulses, with time
$T_R$ between subsequent pulses and time-varying flip angles and phases given by
$\alpha$ and $\phi$. By $\theta$ we denote a length $N_{\rm P}$ vector of local
parameters which include the relaxation times
$T_1$, $T_2$ and the proton density $\rho$.
The discretized spatial domain is represented by 
a rectangular mesh $G_{\rm p}$ of size $m_1 \times m_2$, with grid
spacing 1, defined by
\begin{multline} \label{eq:define_G_p}
G_{\rm p} =
\{ - \lfloor m_1 / 2 \rfloor, \ldots, - \lfloor m_1 / 2 \rfloor + m_1
-1 \} \\
\times
\{ - \lfloor m_2 / 2 \rfloor, \ldots, - \lfloor m_2 / 2 \rfloor + m_2
-1 \} .
\end{multline}
Position on the spatial grid will be denoted by $x = (x_1,x_2)$.

A first approximation to the signal from the macroscopic object,
without thermal noise and spatial discretization effects, is then given by:
\begin{multline} \label{eq:signal_model_with_k_jl}
  s_{j,l} = \sum_{x \in G_{\rm p}} M_j(\theta(x))
  e^{- i k_{j,l} \cdot x } ,\\
  \text{ with $j = 1,\ldots, N_{\rm I}$ and
$l = 1, \ldots, N_{\rm RO}$}
\end{multline}
where $M_j$ is the magnetization at the $j$-th echo, $N_{\rm RO}$ is the total number of read-outs (i.e. snapshots) and  $k_{j,l}$ are the $k$-space sampling locations of the $l$-th sample  during the $j$-th readout interval. Due to our choice of spatial grid,  
$k_{j,l}\in[-\pi,\pi]^2$. Note that $j$ denotes also the snapshot or frame index.
Data is denoted by $d_{j,l}$ and has the same structure as the modeled
signal but may contain thermal noise. The aim of our analysis is to
investigate the interplay between the transient states spin evolution
and the under-sampled $k$-space trajectory. Since we are interested in  heavily under-sampled $k$-space acquisitions, we assume
that the thermal error and the numerical approximation effects of the Discrete Fourier transform are negligible in comparison with aliasing artifacts and thus they will not be taken into account.

The $k$-space data is processed to a
set of snapshot images $I$, defined by
\begin{equation} \label{eq:undersampled_images}
\begin{aligned}
  I(x) = {}& \big( I_1(x), \ldots, I_{N_{\rm I}}(x) \big)
  \\
  I_j(x) = {}& \frac{1}{m_1 m_2}
  \sum_{l} w_{j,l} e^{i k_{j,l} \cdot x}
  d_{j,l}(k) ,
\end{aligned}
\end{equation}
where $x \in G_{\rm p} , j = 1,\ldots,N_{\rm I}$ and $w_{j,l}$ are quadrature or $k$-space density compensation weights. 
By inserting Eq.\ (\ref{eq:signal_model_with_k_jl}) in Eq.
(\ref{eq:undersampled_images}) we obtain
\begin{equation} \label{eq:undersampled_imaged_data}
  I_j(x) = \frac{1}{m_1 m_2}
  \sum_{l} \sum_{y \in G_{\rm p}}
  w_{j,l} e^{i k_{j,l} \cdot (x - y) } M_j(\theta(y)) .
\end{equation}
Based on Eq.\ (\ref{eq:undersampled_imaged_data}), we define
the {\it time-dependent} point spread functions (PSF) associated with the $j$-th snapshot  as
\begin{equation} \label{eq:PSF_per_j}
  P_j(x) = \frac{1}{m_1 m_2} \sum_{l} w_{j,l} e^{i k_{j,l} \cdot x}
\end{equation}
thus Eq.\ (\ref{eq:undersampled_imaged_data}) can be written as a
convolution 
$I_j(x) = \sum_{y \in G_{\rm p}} P_j(x-y) M_j(\theta(y))$.
In the ideally Nyquist sampled $k$-space, each $P_j$ would resemble a
delta function. In MRF this is not the case and $P_j$ causes strong
aliasing artifacts in each snapshot image.

For a typical MRF sequence, the number of frames $N_{\rm I}$ is much larger than the undersampling factor and the $k$-space sampling is varied at each readout to achieve a  full coverage of the spatial frequencies over the whole set of acquisitions. In other words, the {\it average} PSF defined  as
\begin{equation} \label{eq:define_P_from_Pj}
  P(x) = \frac{1}{N_{\rm I}} \sum_{j=1}^{N_{\rm I}} P_j(x) 
\end{equation}
describes a Kronecker delta for realistic MRF experiments and thus it can be considered an alias-free point spread function.

We define the {\it undersampling errors} by
\begin{equation}
  e_{{\rm US},j}(x)
  = I_j(x) - P \ast M_j(\theta(\cdot))(x)
\end{equation}
where $\ast$ denotes convolution. Equivalently, we have:
\begin{equation}
  I_j(x) = P \ast M_j(\theta(x)) + e_{{\rm US},j}(x).
\end{equation}
In MRF,  for each $x$ the
undersampling errors
$e_{\rm US}(x) = \big( e_{{\rm US},j}(x) \big)_{j=1,\ldots,N_{\rm I}}$
are treated as i.i.d. normally distributed noise. This has implications in the sequence design.  For example, in
\cite{MaEtAl2013_MagneticResonanceFingerprinting} it is argued that a
certain amount of randomness in the choice of sequences (e.g.
randomly varying $T_{\rm R}(j)$ and small random variations in
$\alpha(j)$) should guarantee that this assumption is to a large
degree satisfied. On the other hand in \cite{zhao2017optimal,sbrizzi2017dictionary,asslander2017relaxation} it was argued that, taking into account statistical considerations for the least-squares
estimator, optimal sequences are in fact highly structured, that is, flip angle values show clear temporal correlation.
We should emphasize that in general the 
undersampling errors $e_{\rm US}(x)$ are not normally distributed
and do not average out to zero, see also
section 4 of the Supplementary material.

\subsection{Parameter reconstruction}
The MRF  parameter reconstruction is defined as a
least-squares estimator $\theta^*(x)$ of $\theta(x)$:
\begin{equation} \label{eq:MRF_least_squares_1}
  \theta^*(x) =
  \argmin_\theta
  \left\| I(x)  - M(\theta) \right\|^2 .
\end{equation}
To reduce the dimensionality of the minimization problem in Eq.\
(\ref{eq:MRF_least_squares_1}), the fact that $M$
is linear in $\rho$ can be used. Let us denote $\theta = (\eta,\rho)$,
in case $\rho$ is taken as a real parameter, and
$\theta = (\eta , \Re \rho, \Im \rho)$ in case $\theta$ contains the
complex parameter $\rho$.
Using the well known relation between
least-squares estimation and the so called matched filter estimation \cite{DaviesEtAl2014_CS_MRF}, $\theta$
can also be obtained as follows:
\begin{equation} \label{eq:MRF_argmax_1}
  \eta^*(x) =
  \argmax_\eta
  \frac{ | \langle I(x), M(\theta) \rangle | }
  { \| I(x) \| \, \| M(\theta) \| } ,
\end{equation}
where $\langle \cdot,\cdot\rangle$ denotes the usual complex inner
product that is antilinear in the second argument, and
\begin{equation}\label{eq:MRF_argmax_2}
  \rho^*(x) = 
  \frac{ \langle I(x), M(\eta^*(x),1) \rangle }
  { \| M(\eta^*(x),1) \|^2 } .
\end{equation}

In the MR fingerprinting experiments described in the literature, the
maximization as given in Eq.\ (\ref{eq:MRF_argmax_1}) is typically
implemented by using a pre-computed dictionary.

\section{A model for the reconstruction error\label{sec:analysis1}}
To obtain a model for the error, we first derive
the equations that characterize the reconstructed parameters $\theta^*$.  Note that $\theta^*$ is the stationary point of the least-squares
objective function from Eq.\ (\ref{eq:MRF_least_squares_1}).
The corresponding normal equations are nonlinear and are difficult, if not impossible, to solve analytically. Therefore, we will expand $\theta$
and $\theta^*$ as:
\begin{equation} \label{eq:expand_define_theta1}
  \begin{aligned}
    \theta(x) = {}& \theta_0 + \theta_1(x) ,
    \\
    \theta^*(x) = {}& \theta_0 + \theta_1^*(x)
  \end{aligned}
\end{equation}
where $\theta_0$ is a spatially constant value and $\theta_1$, $\theta_1^*$ are the contrast terms in, respectively, the true parameter and the reconstruction. Subsequently, we will linearize $M(\theta)$ and its derivative
$\mathcal{D} M(\theta)$ around $\theta = \theta_0$.
The result will be an equation for $\theta_1^*(x)$, which describes
the errors in MR fingerprinting reconstructions at each spatial
location $x$.

We will show that $\theta_1^*$ can be written in the form
\begin{equation} \label{eq:error_terms_general_behavior}
  \theta_1^*(x) = P \ast \theta_1(x)
  + \epsilon_1(x)
  + \epsilon_2(x, \theta_1(\cdot)) +\text{h.o.t.}
\end{equation}
where $P$ is the point spread function defined
in Eq.\ (\ref{eq:define_P_from_Pj}) and
$\epsilon_1$ and $\epsilon_2$ are error terms, the latter of
  which depends on the {\it function} $\theta_1$. The abbreviation
h.o.t.\ stands for higher order terms in $\theta_1$ and in the derivatives $\mathcal{D}M$. These terms will be discarded in the subsequent  analysis of Eq.\ (\ref{eq:error_terms_general_behavior}).
Note that the term $P \ast \theta_1(x)$ depends purely on
the $k$-space sampling scheme and not on the dynamic behavior of the
magnetization (thus it is independent
on the RF excitation train). Furthermore, the term $\epsilon_1$ is
independent of $\theta_1$, thus this error will in general be present
even in the absence of contrast (homogeneous object).

Crucially, the two error terms
$\epsilon_1(x)$ and $\epsilon_2(x)$  depend on functions
$S_{1;p}^{(1,0)}(x)$, $S_{1;p,q}^{(2,0)}(x)$ and
$S_{1;p,q}^{(1,1)}(x)$. These are convolution kernels which contain information about the time evolution of the magnetization and  $P_j(x)$ and capture the
combined effects of undersampling and transient state sequences.

The rest of this section is dedicated to the derivation of Eq. 
(\ref{eq:error_terms_general_behavior}). To simplify the exposition,
we will start with the case of constant proton density  in
$\theta_0$. Afterward, we
will consider the general case of spatially varying
$\rho_0$.

\subsection{Stationary points of the MRF objective
  functions}

As already mentioned in the previous paragraph, the MRF estimate $\theta^*(x)$ is a stationary point of the objective
function in Eq.\ (\ref{eq:MRF_least_squares_1}). Therefore, 
$\theta^*(x)$ must satisfy the equations
\begin{equation}
  0 = \Re \left\langle M(\theta^*) - I(x) ,
    \pdpd{ M}{\theta_p}(\theta^*)
  \right\rangle ,\,\forall x,\forall p = 1, \ldots, N_{\rm P}
\end{equation}
for $p = 1, \ldots, N_{\rm P}$.
From equations (\ref{eq:undersampled_imaged_data}) and (\ref{eq:PSF_per_j}) it follows that
\begin{multline} \label{eq:stationarity_with_model1}
  0 = \Re \sum_{j=1}^{N_{\rm I}} 
  \overline{ \mathcal{D} M(\theta^*(x))_{j;p}  } M(\theta^*(x))_j\\
  - \Re \sum_{j=1}^{N_{\rm I}} \sum_{y \in G_{\rm p}} P_j(x-y) 
  \overline{ \mathcal{D}M(\theta^*(x))_{j;p} } M(\theta(y))_j    
\end{multline}
where  $\mathcal{D} M(\theta)$ denotes the jacobian matrix of $M$, $\mathcal{D}M(\theta)_{j;p}$ are its components and the overscoring indicates complex conjugation.

\subsection{Expansion of the terms in Eq.\
  (\ref{eq:stationarity_with_model1})}
\label{subsec:expansion_stationarity_constant_rho0}

The next step is to replace $\theta(x)$ and $\theta^*(x)$ by
$\theta_0+\theta_1(x)$ and $\theta_0+\theta_1^*(x)$ and expand
Eq.\ (\ref{eq:stationarity_with_model1}) to first order
in $\theta_1(x)$ and $\theta_1^*(x)$.
Using first order Taylor expansions for $M(\theta)$ and $\mathcal{D}M(\theta)$ around $\theta_0$, we straightforwardly obtain 
\begin{multline}
  \overline{ \mathcal{D} M(\theta^*(x))_{j;p} } M(\theta(y))_j
  =  
  \overline{ \mathcal{D}M(\theta_0)_{j;p} } M(\theta_0)_j\\
  + \sum_{q=1}^{N_P}
  \overline{ \mathcal{D}^2 M(\theta_0)_{j;p,q} } M(\theta_0)_j  \theta_{1,q}^*(x)
  \\
   + \sum_{q=1}^{N_P} \overline{ \mathcal{D}M(\theta_0)_{j;p} }
  \mathcal{D} M(\theta_0)_{j;q} \theta_{1,q}(y)  + \text{h.o.t.}
\end{multline}
Defining 
\begin{equation} \label{eq:define_SU10_SU20_SU11}
  \begin{aligned}
    S_{p,q}^{(1,1)}(x) = {}&
    \sum_{j=1}^{N_I} P_j(x)
    \overline{ \mathcal{D}M(\theta_0)_{j;p} } \mathcal{D} M(\theta_0)_{j;q} 
    \\
    S_{p}^{(1,0)}(x) = {}&
    \sum_{j=1}^{N_I} P_j(x)
    \overline{ \mathcal{D} M(\theta_0)_{j;p} } M(\theta_0)_j
    \\
    S_{p,q}^{(2,0)}(x) = {}&
    \sum_{j=1}^{N_I} P_j(x)
    \overline{ \mathcal{D}^2 M(\theta_0)_{j;p,q} } M(\theta_0)_j
\end{aligned}
\end{equation}
we observe that, to first order, the second term in
Eq.\ (\ref{eq:stationarity_with_model1}) may be written as
\begin{equation}
  -\Re \bigg[ S_{p}^{(1,0)} \ast 1 (x)
  + \sum_{q=1}^{N_P} S_{p,q}^{(1,1)} \ast \theta_{1,q}(x) 
  + \sum_{q=1}^{N_P} \theta_{1,q}^* S_{p,q}^{(2,0)} \ast 1(x) \bigg]
\end{equation}
 where $1(x)$ denotes the constant function with value 1 at all locations $x$
in the mesh.

The sums defined in Eqs.\
(\ref{eq:define_SU10_SU20_SU11}) are an essential
element of the analysis. They are weighted sums of the (time-dependent) point spread
functions, with the ``weights'' given by linear-antilinear terms
$\overline{ \mathcal{D} M(\theta_0)_{j;p} } M(\theta_0)_j$ etc.  The sums contain
the combined effects of $k$-space undersampling and time-dependent behavior of the magnetization.

We further split the terms in Eq.\ (\ref{eq:stationarity_with_model1}) into ``mean'' and ``residual'' parts. We therefore define
\begin{equation} \label{eq:define_S0_11_and_S1_11}
  \begin{aligned}
    S_{\text{mean};p,q}^{(1,1)}(x) = {}&
    P(x) \sum_{j=1}^{N_I}
    \overline{ \mathcal{D}M(\theta_0)_{j;p} } \mathcal{D} M(\theta_0)_{j;q} 
    \\
    S_{\text{resid};p,q}^{(1,1)}(x) = {}&
    \sum_{j=1}^{N_I} ( P_j(x) - P(x) )
    \overline{ \mathcal{D}M(\theta_0)_{j;p} } \mathcal{D} M(\theta_0)_{j;q} 
  \end{aligned}
\end{equation}
such that
\begin{equation} \label{eq:sum_S0_11_and_S1_11}
  S_{p,q}^{(1,1)}(x)  = S_{\text{mean};p,q}^{(1,1)}(x) + S_{\text{resid};p,q}^{(1,1)}(x) .
\end{equation}
This decomposition separates the effects of time-varying (residual)
and constant (mean) sampling. It will turn out that
the error terms are proportional to the residual parts.
Consistently with this, the residual component vanishes when
there is no undersampling (i.e. $P_j = P$) and/or when the
magnetization is in the steady states, which results into the
``weights'' $\overline{ \mathcal{D}M(\theta_0)_{j;p} } \mathcal{D}
M(\theta_0)_{j;q}$ being  time-independent. This is the case for
conventional MRI acquisitions.

Similarly, we define $S_{\text{mean};p}^{(1,0)}(x)$, $S_{\text{resid};p}^{(1,0)}(x)$
and $S_{\text{mean};p,q}^{(2,0)}(x)$, $S_{\text{resid};p,q}^{(2,0)}(x)$
replacing the weights in Eq.\ (\ref{eq:define_S0_11_and_S1_11})
by the weights used in defining $S_{p}^{(1,0)}(x)$ and $S_{p,q}^{(2,0)}(x)$, respectively.

We proceed with the first term in 
Eq.\ (\ref{eq:stationarity_with_model1}). Using again the
Taylor expansions of $M(\theta)$ and
$\mathcal{D} M(\theta)$ and the definition of mean and residual components, 
this term can be written as
\begin{multline}
  \Re \bigg[ S_{\text{mean};p}^{(1,0)} \ast 1(x)
  + \sum_{q=1}^{N_P} \theta_{1,q}^*S_{\text{mean};p,q}^{(1,1)} \ast 1(x) \\
  + \sum_{q=1}^{N_P} \theta_{1,q}^* S_{\text{mean};p,q}^{(2,0)} \ast 1(x) \bigg] .
\end{multline}

Finally, in our expansion of
Eq.\ (\ref{eq:stationarity_with_model1}),
Eq.\ (\ref{eq:sum_S0_11_and_S1_11}) is used (and the similar
property for $S_{p}^{(1,0)}(x)$ and $S_{p,q}^{(2,0)}(x)$)
to obtain some cancellations, and make a clear identification of
error terms possible.
The first order expansion of Eq.\ 
(\ref{eq:stationarity_with_model1}) is thus:
\begin{multline} \label{eq:expanded_stationarity_constant_rho0}
    0 = \Re \bigg[ 
    \sum_{q=1}^{N_P} \theta_{1,q}^* S_{\text{mean};p,q}^{(1,1)} \ast 1
    -  \sum_{q=1}^{N_P} S_{\text{mean};p,q}^{(1,1)} \ast \theta_{1,q}
    \\ - S_{\text{resid};p}^{(1,0)} \ast 1 
    - \sum_{q=1}^{N_P} S_{\text{resid};p,q}^{(1,1)} \ast \theta_{1,q}\\ 
    - \sum_{q=1}^{N_P} \theta_{1,q}^*  S_{\text{resid};p,q}^{(2,0)} \ast 1 \bigg] .
\end{multline}
This was obtained as the terms $\pm S_{\text{mean};p}^{(1,0)} \ast 1 (x)$ and $\pm S^{(2,0)}_{\rm mean}\ast 1 (x)$
cancel each other. The above equation is a formal expansion of Eq.\
(\ref{eq:stationarity_with_model1})  in the variables
$\theta_1$, $\theta_1^*$,
$S_{\text{resid};p}^{(1,0)}$, $S_{\text{resid};p,q}^{(2,0)}$ and $S_{\text{resid};p,q}^{(1,1)}$.

\subsection{Error model for constant proton density reference}
\label{subsec:linear_system_theta1}

Equation (\ref{eq:expanded_stationarity_constant_rho0})
is a linear system for the MRF estimate $\theta_1^*(x)$ for each $x$. As a next step, we write down the
solution of this system and identify the correct contribution and the
systematic errors as outlined in Eq.\ (\ref{eq:error_terms_general_behavior}).\\
We first observe that the function
$S_{\text{mean};p,q}^{(1,1)}(x)$ is simply given by $P(x) N_{p,q}$ where
$N_{p,q}$ is defined as
\begin{equation}
  N_{p,q} = \sum_{j=1}^{N_{\rm I}}
  \overline{\mathcal{D} M(\theta_0)_{j;p}}
  \mathcal{D} M(\theta_0)_{j;q} .
\end{equation}
The PSF $P$ is well-behaved, that is, it approaches a Kronecker delta, thus $P \ast 1 \approx 1$. As a consequence,
the first and second term in (\ref{eq:expanded_stationarity_constant_rho0})
can be approximated by, respectively,
$\sum_q \Re N_{p,q} \theta^*_{1,q}(x)$ and
$\sum_q \Re N_{p,q} ( P \ast \theta_{1,q}(x) )$.

The fifth term in
  (\ref{eq:expanded_stationarity_constant_rho0}) is a product of 
two factors assumed to be small, namely of $\theta_1^*$ and of
$S_{\text{resid};p,q}^{(2,0)}$. Staying with our philosophy of
keeping only the first order terms, we will omit it.
Defining two vector valued functions 
\begin{equation} \label{eq:define_errors_E1_E2_E3}
  \begin{aligned}
    E_{1;p}(x) = {}& \Re S_{\text{resid};p}^{(1,0)} \ast 1 (x)
    \\
    E_{2;p}(x) = {}& \Re \sum_q S_{\text{resid};p,q}^{(1,1)} \ast \theta_{1,q}(x)  ,
  \end{aligned}
\end{equation}
we conclude that $\theta_1^*(x)$ is given to first order approximation
by
\begin{multline} \label{eq:theta1_with_errors}
  \theta_1^*(x) = P \ast \theta_1(x) +
  ( \Re N )^{-1}\left(E_1(x) + E_2(x) \right) ,
\end{multline}
where it was used that $P$ is real.
The first term on the right hand side is identified as the correct
image.
The other two terms in Eq.\ (\ref{eq:error_terms_general_behavior})
are given by
$\epsilon_j = ( \Re N )^{-1}  E_j$, $j=1,2$.

\subsection{Error model for variable proton density reference}
\label{subsec:var_ref_pd}

So far, we have assumed that all components of $\theta$ are close to
some constant reference value. In
Section 1
of the Supplementary material a more refined model is derived that allows for a variable proton density. We briefly explain why this is of interest. Firstly, the proton density is always zero outside the object (air)
and therefore can hence hardly be considered ``nearly
constant''. The second reason is better illustrated in the $k$-space domain. Note that by taking the Fourier transform on both sides of the error $E_1$ defined in Eq.\ (\ref{eq:define_errors_E1_E2_E3}) we obtain:
\begin{equation*}
\widehat{E}_{1;p}(k) = \mathcal{F}\left\{ \text{Re} S_{\text{resid};p}^{(1,0)}\ast 1\right\} (k).
\end{equation*}
where $\widehat{E}_{1;p}$ denotes the Fourier transform of $E_{1;p}$.
Convolution by the constant function 1 becomes a product in the $k$-space with a Dirac delta centered at $k=0$, thus the previous expression vanishes for all $k\neq0$. For $k=0$ we obtain:
\begin{equation}
  \widehat{S}_{\text{resid};p}^{(1,0)}(0) = 
  \sum_{j=1}^{N_{\rm I}} ( \widehat{P}_j(0) - \widehat{P}(0) )
  \overline{ \mathcal{D}M(\theta_0)_{j;p} } M(\theta_0)_{j}
\end{equation}
where $\widehat{P}_j(k)$ and $\widehat{S}_{\text{resid};p}^{(1,0)}(k)$ denote the Fourier transforms of $P_j(x)$ and $S_{\text{resid};p}^{(1,0)} $, respectively.

In case of radial or spiral sampling, the $k=0$ Fourier component
is sampled at each interval, so that $\widehat{P}_j(0)$ is independent
of $j$ and equal to $\widehat{P}(0)$. Therefore, for radial or spiral
sampling, $E_{1;p}(x)$ effectively vanishes. By allowing for a
variable proton density a better approximation for this type of error
is found that does not vanish.

In case of variable reference proton density 
the equivalent of Eq.\ (\ref{eq:theta1_with_errors})
is given by (see the Supplementary material)
\begin{empheq}{align} \label{eq:theta1_with_errors_varrho0}
  \theta_1^*(x) &=
  |\rho_0^*(x) |^{-2} ( \Re N )^{-1} \times \nonumber\\
  {}&\left( \Re \overline{\rho_0^*(x)}
    N  P \ast (\rho_0 \theta_1)(x) + E_1(x) + E_2(x) \right),
\end{empheq}
where $\rho_0^*(x)  = P \ast \rho_0(x)$
and
\begin{equation} \label{eq:define_errorsE_varrho0}
  \begin{aligned}
    E_{1;p}(x) = {}&
    \Re \overline{\rho_0^*(x)} S_{\text{resid};p}^{(1,0)} \ast \rho_0 (x)
    \\
    E_{2;p}(x) = {}&
    \Re \overline{\rho_0^*(x)} \sum_q S_{\text{resid};p,q}^{(1,1)} \ast (\rho_0 \theta_{1,q})(x).  
  \end{aligned}
\end{equation}

In regions where $\rho_0$ varies, $\theta_1^*$ is in general no longer equal to $P \ast \theta_1$.
However, inside the object we typically choose $\rho_0$ constant, so that $\rho_0^*$ equals $\rho_0$ and the first term reduces 
again to $P \ast \theta_1$, which we assume is small. In other words, $\rho_0$ is a binary valued function (i.e. a mask) which attains 0 in the locations outside the brain (air)).
Note that, in this case, the error terms of the generalized error model of Eq.  (\ref{eq:error_terms_general_behavior}) are given by
$\epsilon_j(x) = |\rho_0^*(x) |^{-2} ( \Re N )^{-1} E_j(x)$ and they
satisfy the properties introduced just after Eq.\ (\ref{eq:error_terms_general_behavior}).

\section{Model validation}\label{sec:tests}
In this section, we will investigate the validity of Eq. 
(\ref{eq:theta1_with_errors_varrho0}) as an MRF error model by means of numerical examples. In particular,  we will consider standard, well-established MRF acquisition schemes and we will show that the error predicted by
Eq.\ (\ref{eq:theta1_with_errors_varrho0}) is indeed a good approximation of the error obtained by actual MRF reconstructions.
Subsequently, in section \ref{sec:in-depth_analysis} we will leverage on our model to investigate and uncover different aspects of the MRF  paradigm.

The examples that follow focus on 2D gradient spoiled sequences with radial, spiral and Cartesian $k$-space samplings where 
$\theta = ( \log T_1, \log T_2,\rho)$. The logarithmic change of
variable is meant to scale the relaxation times to a similar range. Other quantities that may be important are, for example, the
relaxation time $T_2^*$ in presence of intravoxel dephasing and
the transmit source field $B_1^+$. To keep our analysis within
practical constraints, we do not consider them.

\subsection{General simulation setup}
\label{subsec:general_setup}
To model the spoiling gradient effects, we compute the macroscopic
voxel signal as a sum of differently resonating isochromat responses,
each of which is modeled using the Bloch equations. Additional phase
accrual effects caused by off-resonance are not taken into account
since we focus on gradient spoiled sequences. Detailed information
regarding the signal simulations and image reconstructions is reported
in
Section 2
of the Supplementary material.

\begin{figure}
\begin{center}
    \includegraphics[width=7cm]{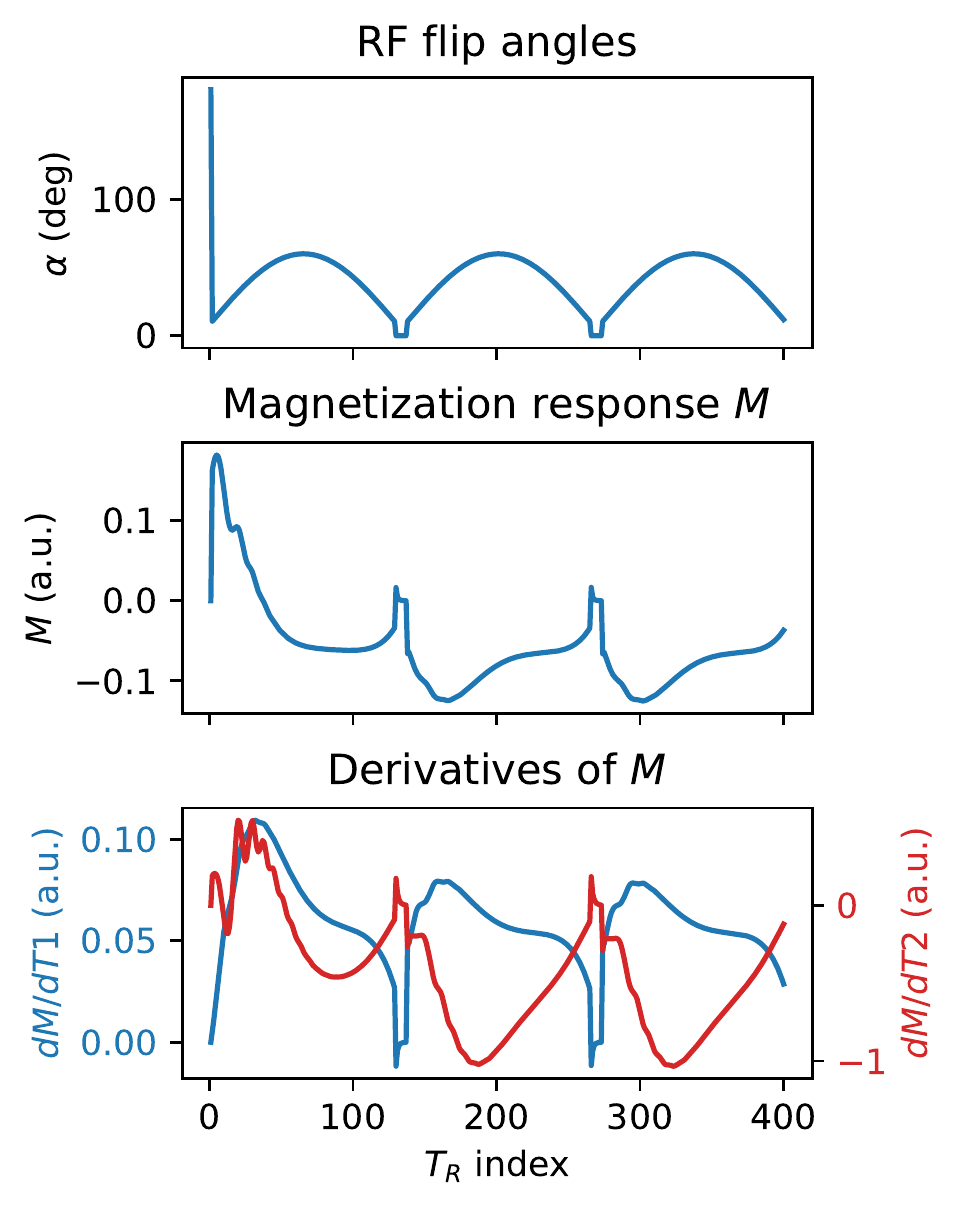}
\end{center}
  \caption{Flip angles and plot of $M(t,\theta)$ and some derivatives
    for $T_1 = 1.0$ s, $T_2 = 0.08$ s, PD = 1 [a.u.], for sequence 1.} 
\label{fig:SeqInfo_Seq1}
\end{figure}

As it is common in MRF, an inversion pulse precedes a time dependent
flip angle train. The flip angles  vary between 0 and $60^{\text{o}}$
and have a $90^{\text{o}}$  phase with respect to the inversion pulse,
see also Fig. \ref{fig:SeqInfo_Seq1}. Although our analysis holds for
any choice of echo and repetition times ($T_{\rm E}$ and $T_{\rm R}$),
including temporally varying values, in this work we consider only
fixed (time independent) values: $T_{\rm R}=15$ ms and $T_{\rm E} =
7.5$ ms. This sequence will be referred to as sequence 1.
In this section three sampling schemes are employed: radial golden angle, spiral golden
angle and  Cartesian. The resolution is $128\times 128$ voxels and the undersampling factors for each image
are 32, 32 and 16, respectively. The undersampling factors are defined in the angular, radial and phase encoding direction, respectively. We opt for a milder undersampling factor in the Cartesian acquisition since, as we will show, this kind of sequences is more susceptible to undersampling artifacts; a factor of 32 would lead to impracticable results.

MR fingerprinting reconstructions are performed by solving Eq. (\ref{eq:MRF_argmax_1}) using a precomputed dictionary of complex
signal evolutions $M(T_1,T_2,1)$. Here $T_1,T_2$ are chosen in logarithmic mesh
with grid distances approximately $0.5 \%$ for $T_1$ and approximately
$1 \%$ for $T_2$. With double precision computations
this lead to a dictionary of a manageable size of about 1GB, and to an
accuracy that is sufficient to compare modeled and MRF errors in the subsequent analysis.

Error predictions according to our model are obtained by numerical
solution of Eq.\ (\ref{eq:theta1_with_errors_varrho0}).
The first term in Eq.\ (\ref{eq:theta1_with_errors_varrho0}) is
defined to be the correct solution.
Having specified the function $\rho_0(x)$, it is straightforward to
compute all the quantities in
(\ref{eq:define_errorsE_varrho0})
and to solve the matrix
equation for $\theta_1^*(x)$ for all $x$ in the FOV.
The computations are implemented in the Julia programming language \cite{bezanson2017julia}. The convolutions make use of 
NUFFT while the derivatives involved in the $S^{(\alpha,\beta)}$ are evaluated by automatic differentiation. 
In the numerical solution of (\ref {eq:theta1_with_errors_varrho0}-\ref{eq:define_errorsE_varrho0}), the most computationally intensive steps are the convolutions involving spatially dependent quantities. These are done by applying forward and adjoint NUFFTs for each index $j$. The remaining steps are relatively cheap. The whole process takes about 3 minutes on a 16 cpu linux machine.

\subsection{Test 1.1: checkerboard phantom}
As a first test, a checkerboard model with variations of
$\pm 25 \%$ in $T_1$ and smaller variations in $T_2$ is considered.
The parameter values
$(T_1,T_2) = (750,70)$ ms and $(T_1,T_2) = (1250,90)$ ms
roughly correspond to typical white
and gray matter values, respectively. The reference proton density,
$\rho_0$, is chosen to be equal to the true proton density that is, 1 inside the checkerboard and 0 outside.

Figure~\ref{fig:errormodel_checkerboard_1} shows the validation results for this phantom and the three  sampling schemes. 
Root mean squared (RMS) averages of the actual MRF errors, the predicted errors and the partial error contributions $\epsilon_1(x)$ and $\epsilon_2(x)$  are given in Table~\ref{tab:rmsErr}.
Further examples of checkerboard phantoms for larger parameter
variations  are given in
section~3
of the Supplementary material.

\subsection{Test 1.2: numerical brain model}
\label{subsec:brain_phantom}
The second example concerns a numerical brain phantom \cite{kwan1999mri}.  In this case $\rho_0$ is
chosen equal to $0.8$ a.u. inside the head and zero otherwise. The acquisition and reconstruction setups are  the
same as in the previous test. The results are displayed in
Figure~\ref{fig:errormodel_brain_1} and summarized in
Table~\ref{tab:rmsErr}.

From these two validation tests we observe the following.\\
(a) When parameters vary moderately 
 (e.g.\ $\pm 25 \%$ compared to the reference value, or a contrast of about a factor 1.5) inside the FOV, the predicted  imaging errors according to our model from Eq.\ (\ref{eq:theta1_with_errors_varrho0}) are in close agreement with
the observed MRF imaging errors.\\
(b) When much larger parameter variations are present,
some degradations occur, particularly in regions of extremely small or
large parameters.
Nonetheless, the overall error estimation is still qualitatively
similar, and predicted and actually obtained
errors are of the same order of magnitude.
Therefore, Eq.\ (\ref{eq:theta1_with_errors_varrho0}) is still valid as a predictive error model.\\
(c) The Cartesian sampling is clearly a sub-optimal acquisition
scheme and incapable of returning acceptable
parameter maps in this 16-fold acceleration case.

\newlength{\mylength}
\setlength{\mylength}{72mm}
\begin{figure*}
  \begin{center}
    \begin{minipage}[t]{\mylength}
      \centerline{(a) radial}
      \includegraphics[width=72mm]{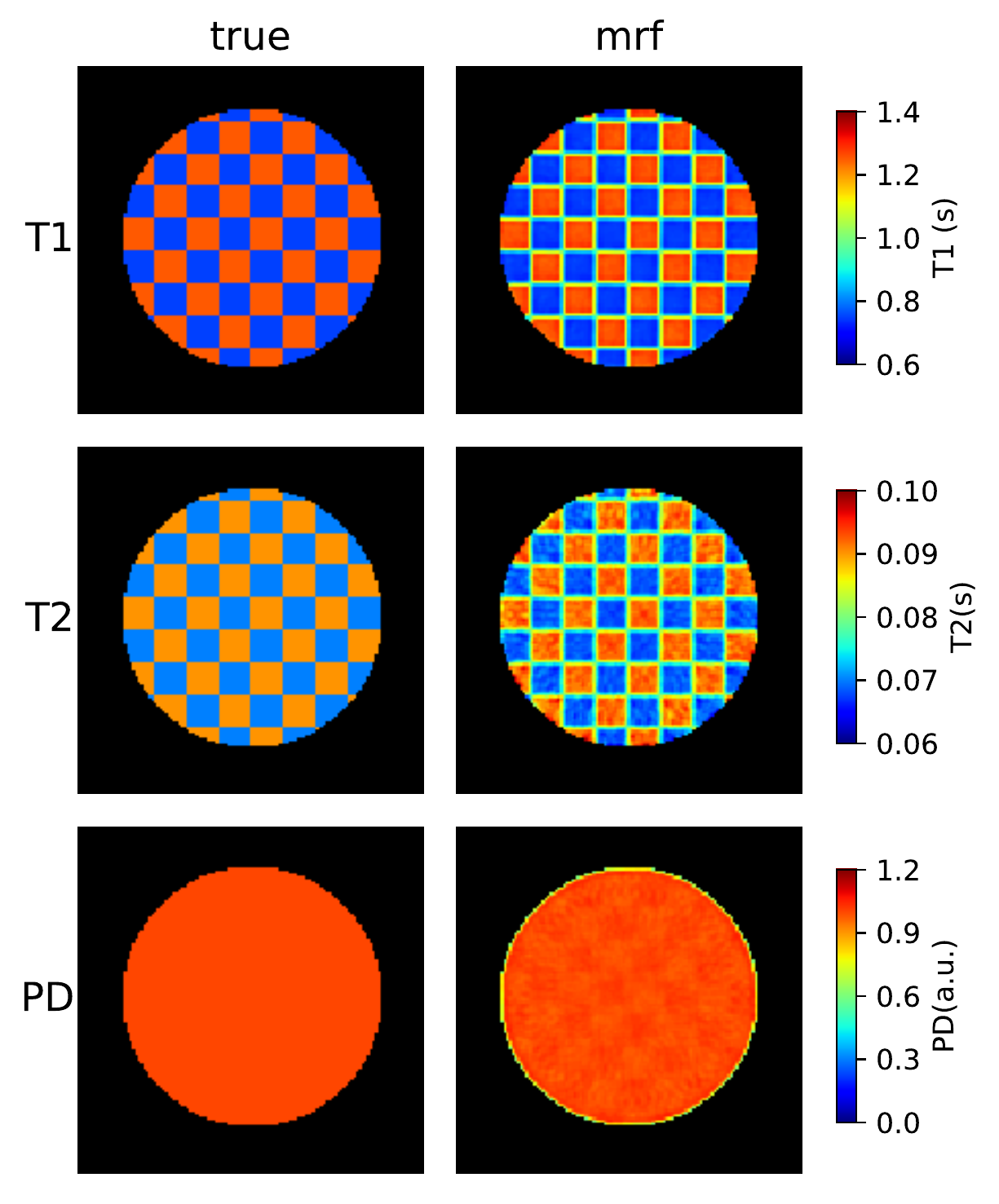}
    \end{minipage}
    \begin{minipage}[t]{\mylength}
      \centerline{(b) radial}
      \includegraphics[width=72mm]{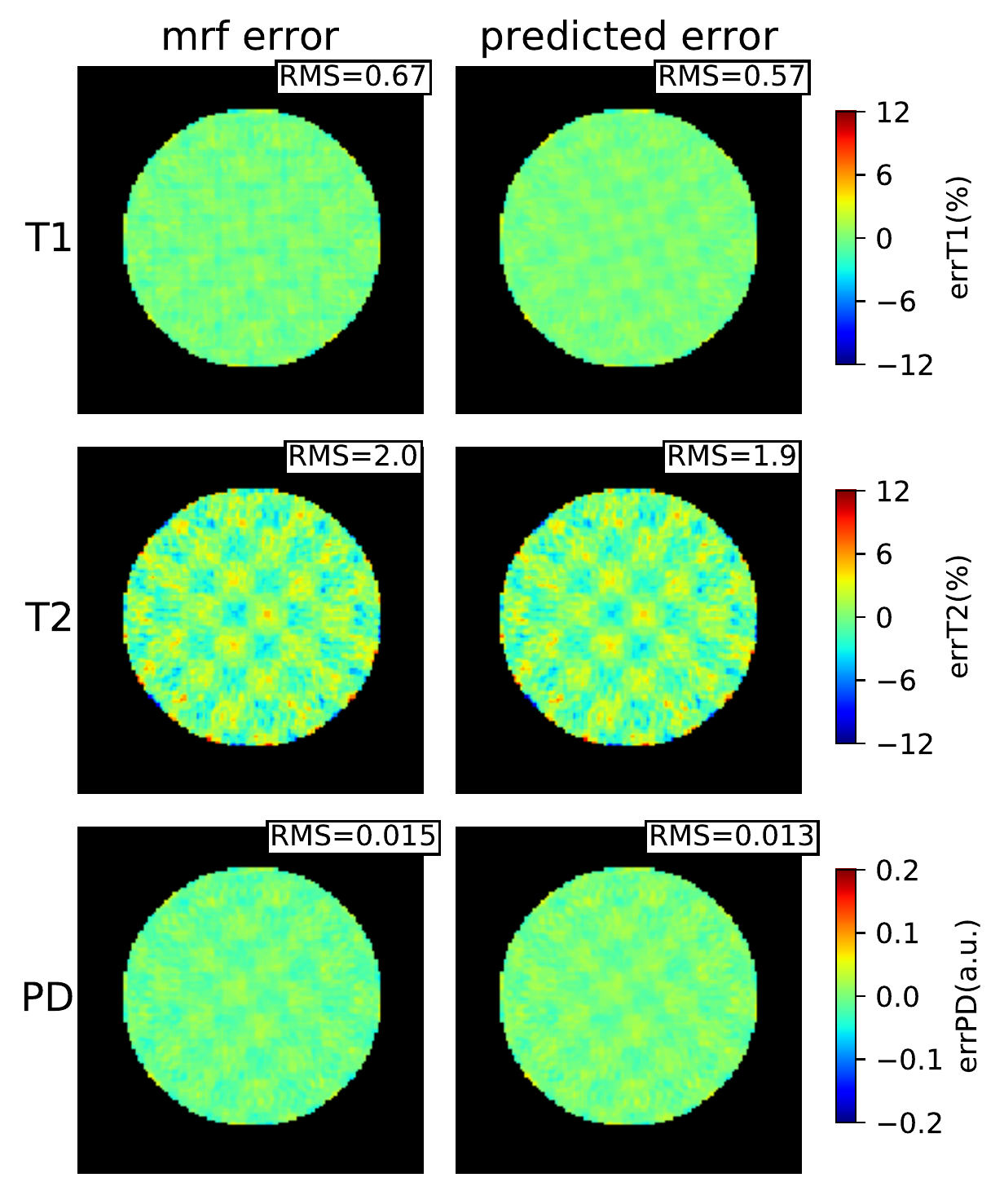}
    \end{minipage}
    \begin{minipage}[t]{\mylength}
      \centerline{(c) spiral}
      \includegraphics[width=72mm]{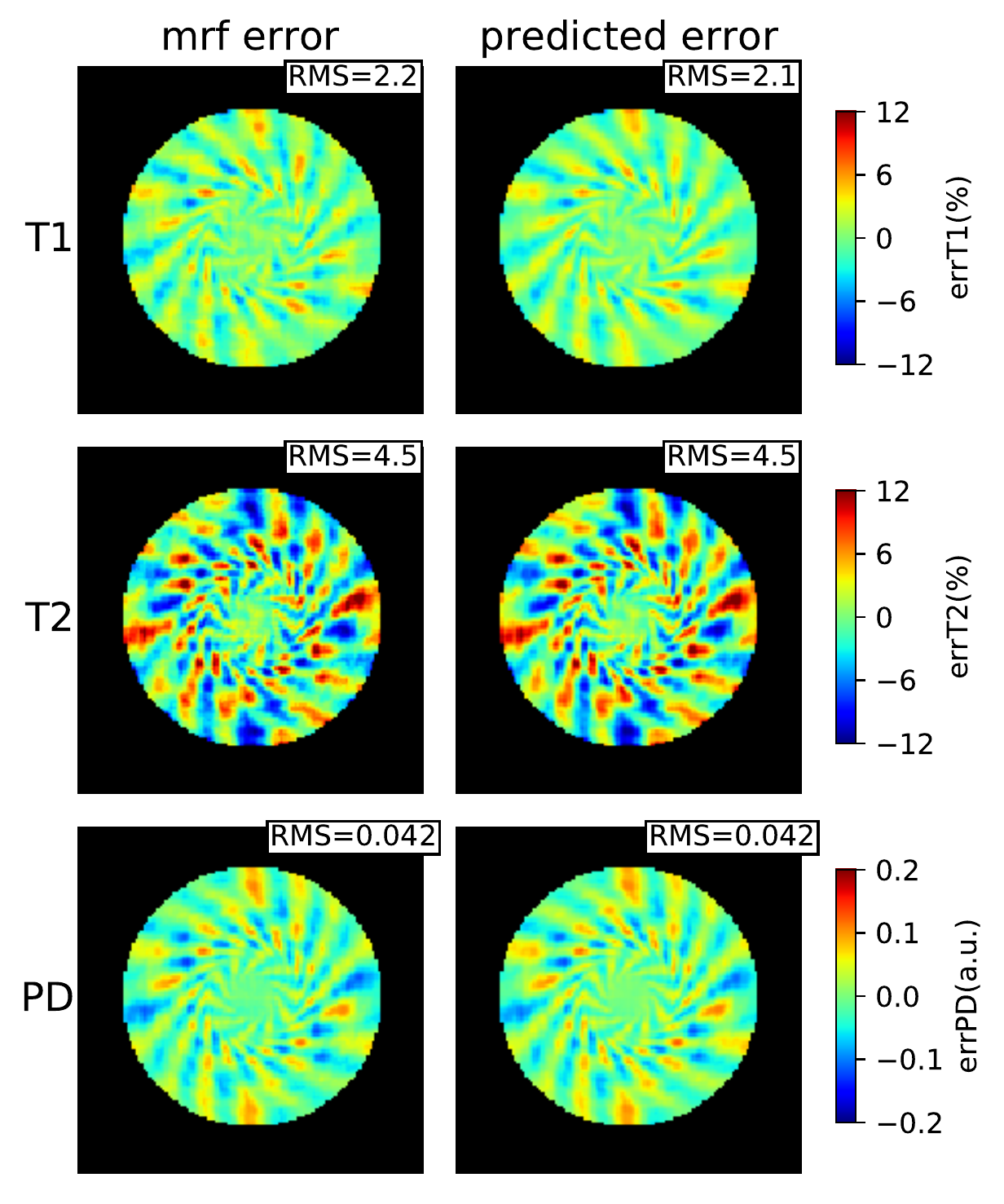}
    \end{minipage}
    \begin{minipage}[t]{\mylength}
      \centerline{(d) Cartesian}
      \includegraphics[width=72mm]{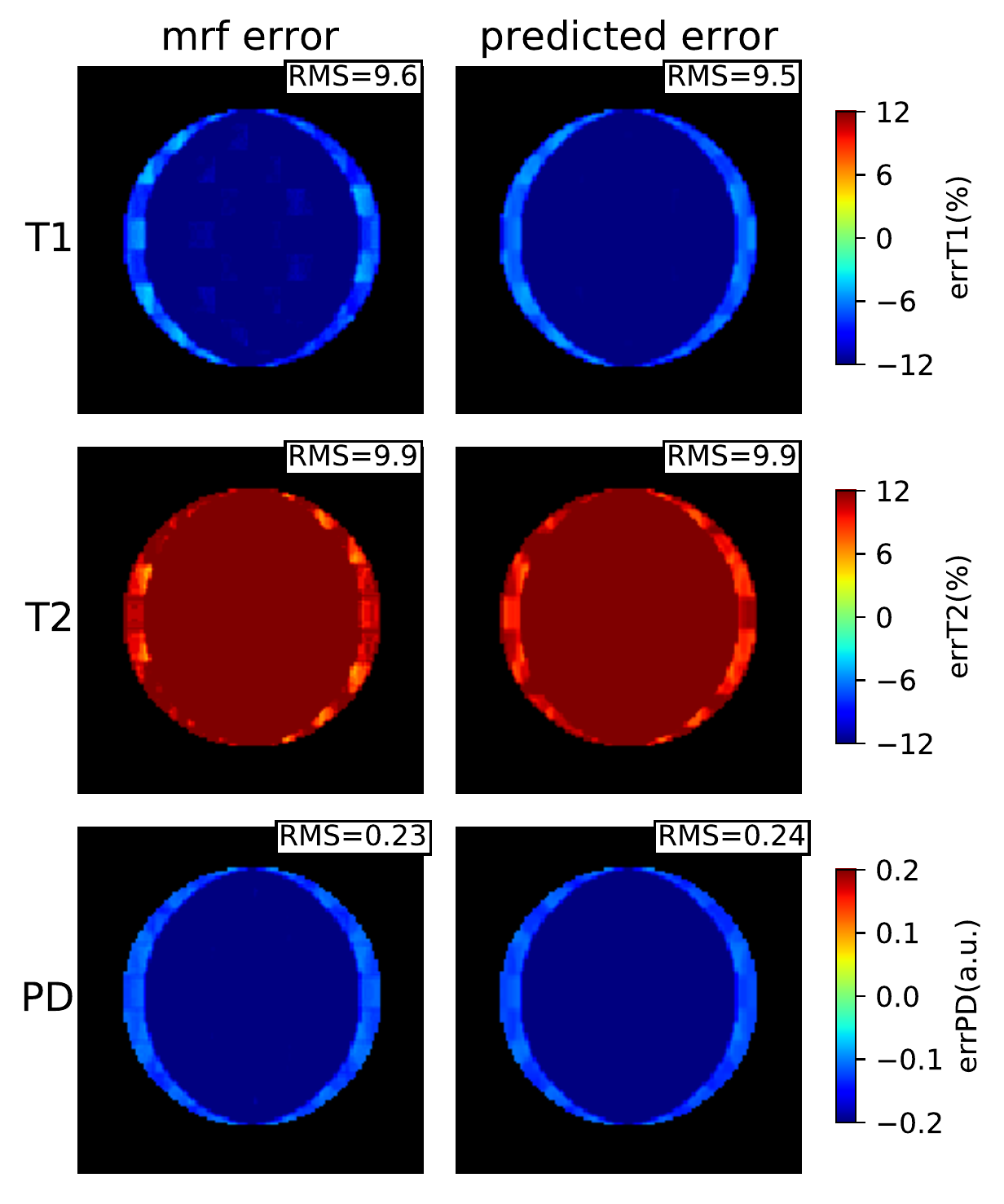}
    \end{minipage}
  \end{center}
  \caption{%
    Test 1.1: A numerical checkerboard phantom.
    (a) true model and a MRF reconstruction for radial sampling;
    (b) actual and predicted MRF errors for radial sampling;
    (c) as (b) for spiral sampling;
    (d) as (b) for Cartesian sampling.
    The rows concern $T_1$, $T_2$ and PD
    respectively.  RMS values are reported in the units
      of the image they refer to.}
    \label{fig:errormodel_checkerboard_1}
\end{figure*}
\begin{figure*}
  \begin{center}
    \begin{minipage}[t]{\mylength}
      \centerline{(a) radial}
      \includegraphics[width=72mm]{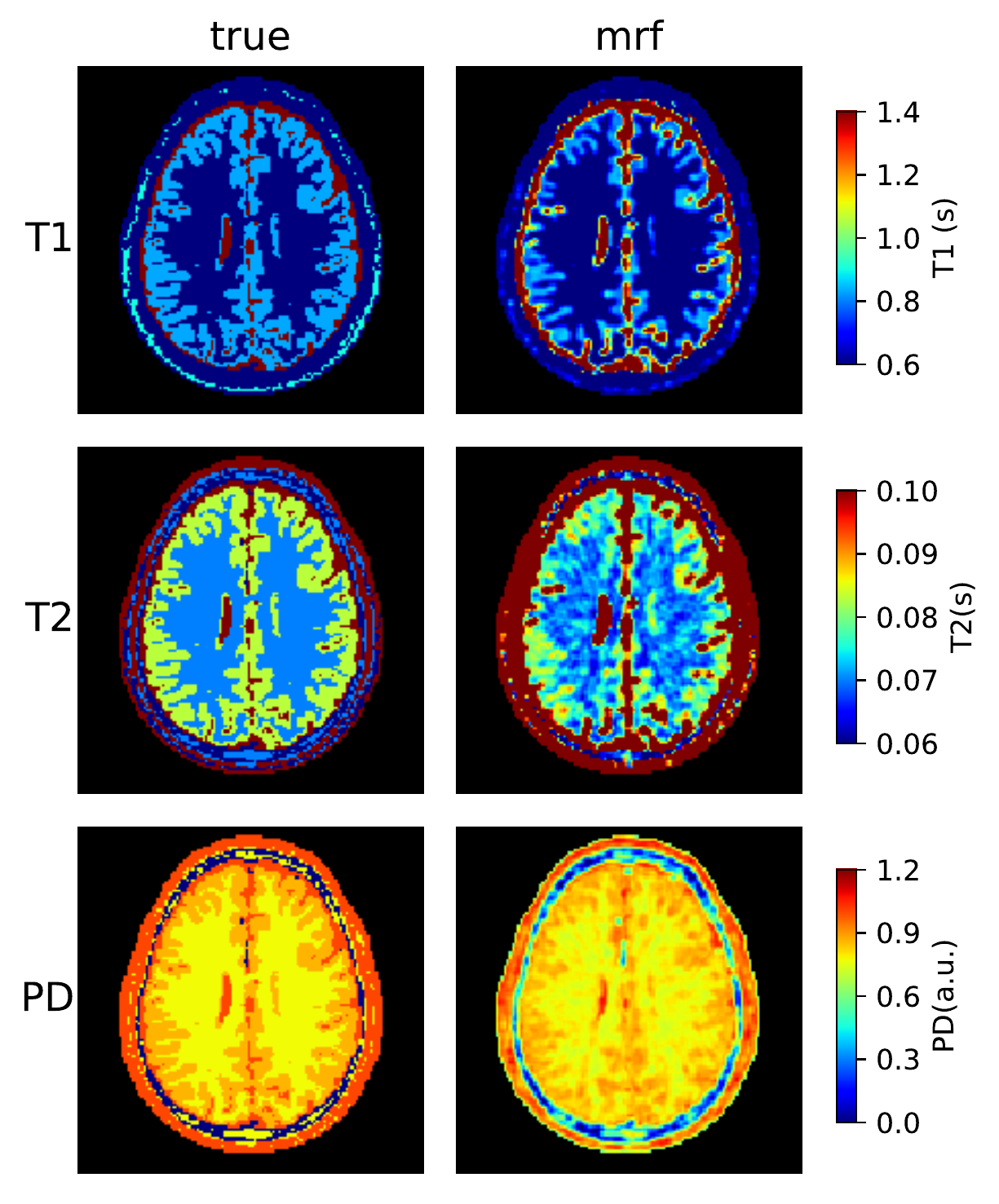}
    \end{minipage}
    \begin{minipage}[t]{\mylength}
      \centerline{ (b) radial}
      \includegraphics[width=72mm]{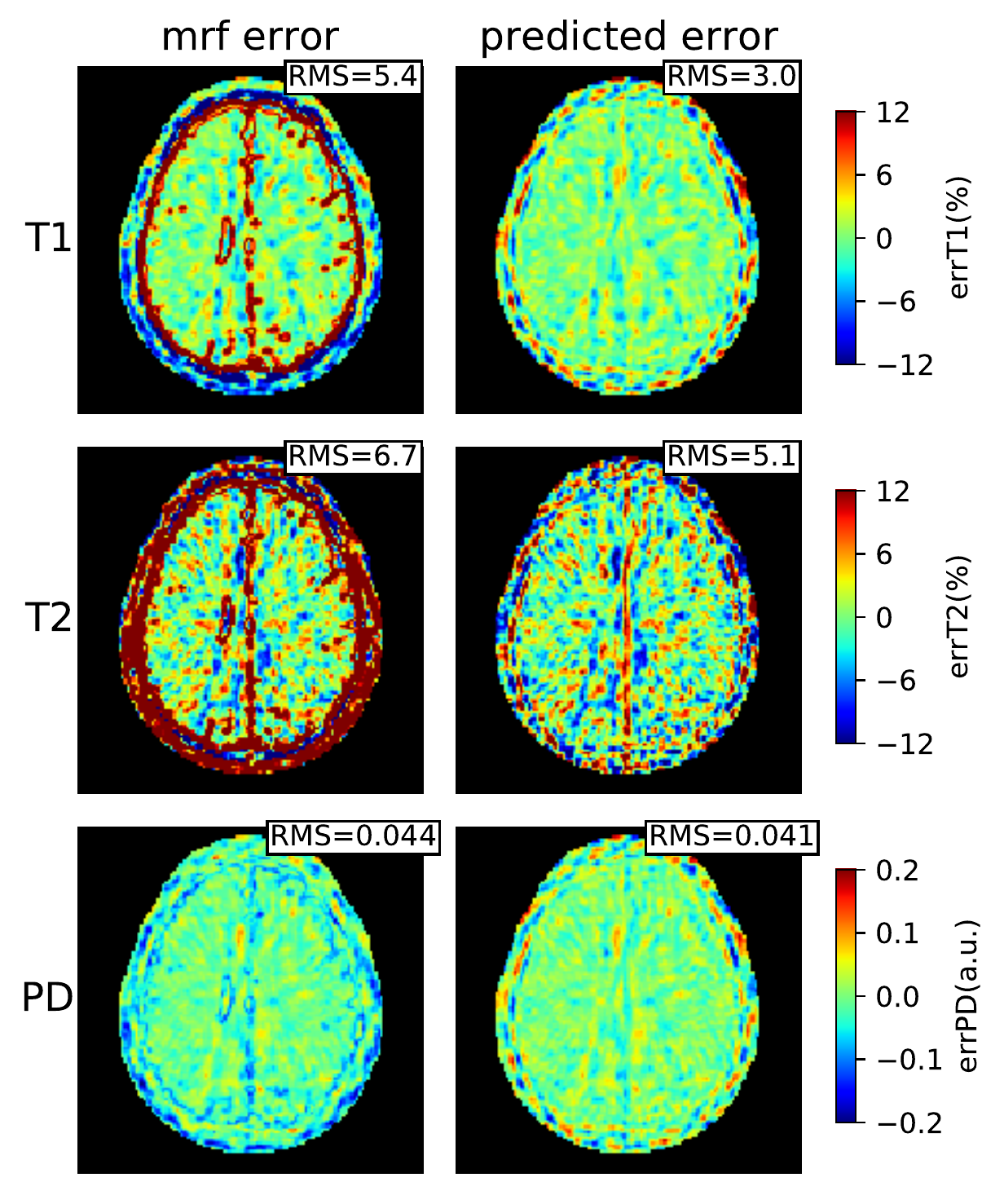}
    \end{minipage}
    \begin{minipage}[t]{\mylength}
      \centerline{ (c) spiral}
      \includegraphics[width=72mm]{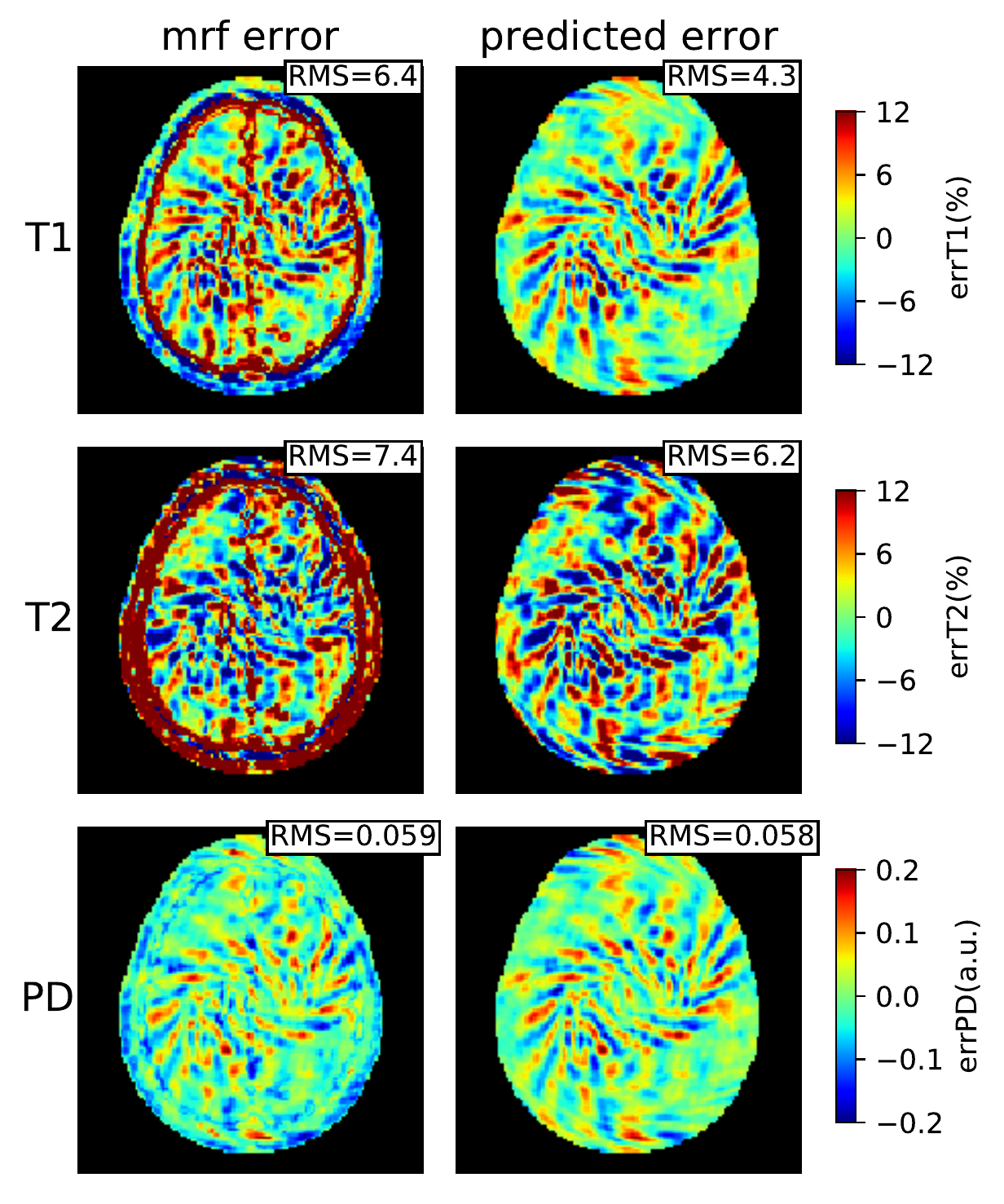}
    \end{minipage}
    \begin{minipage}[t]{\mylength}
      \centerline{ (d) Cartesian}
      \includegraphics[width=72mm]{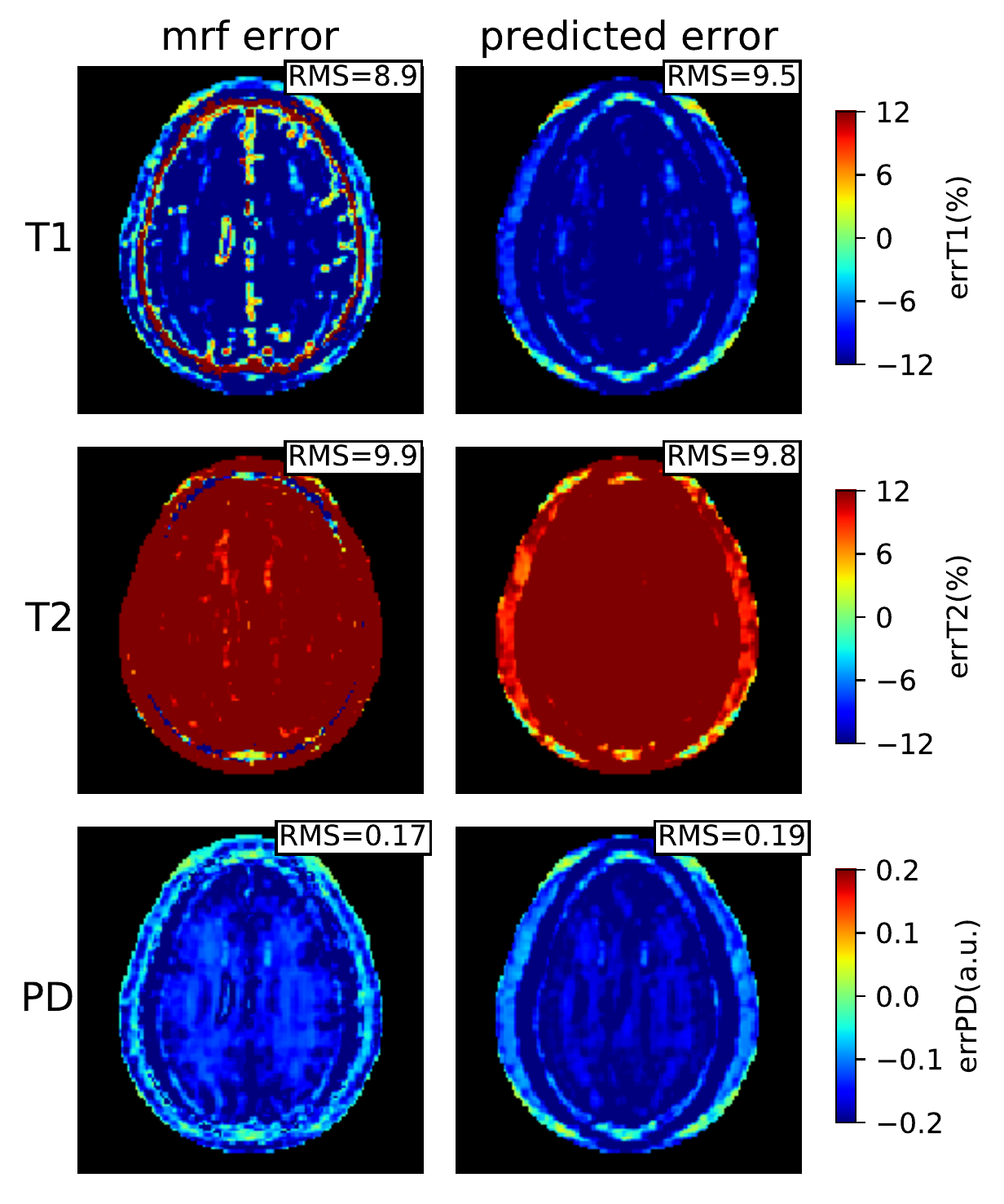}
    \end{minipage}
  \end{center}
  \caption{%
    Test 1.2: A brain phantom. True model, MRF reconstruction,
  predicted and observed errors as in
  Figure~\ref{fig:errormodel_checkerboard_1}.  RMS values are reported in the units
      of the image they refer to.}
    \label{fig:errormodel_brain_1}
\end{figure*}

\begin{table*}\centering
\caption{Predicted error values}
  \label{tab:rmsErr}
  \begin{tabular}{|l|lll|lll|lll|lll|lll|} \hline
    & \multicolumn{15}{c|}{{\bf Test 1.1 (checkerboard)}}
    \\  \hline                                           
    & \multicolumn{3}{c|}{radial, seq1}
    & \multicolumn{3}{c|}{spiral, seq1}
    & \multicolumn{3}{c|}{Cart.\, seq1}
    & \multicolumn{3}{c|}{random, seq1}
    & \multicolumn{3}{c|}{radial, seq2}
    \\
    & err & $\epsilon_1$ & $\epsilon_2$ 
    & err & $\epsilon_1$ & $\epsilon_2$ 
    & err & $\epsilon_1$ & $\epsilon_2$ 
    & err & $\epsilon_1$ & $\epsilon_2$ 
    & err & $\epsilon_1$ & $\epsilon_2$ 
    \\ \hline
    T1 (\%)
    & 0.6 & 0.4 & 0.4
    & 2.1 & 2.0 & 0.6
    &  12 &  12 & 0.5
    & 4.0 & 1.1 & 3.9
    & 5.7 & 4.7 & 3.2
    \\
    T2 (\%)
    & 1.9 & 1.1 & 1.6
    & 4.6 & 4.4 & 1.3
    &  21 &  21 & 0.9
    & 7.3 & 4.2 & 5.9
    & 6.9 & 5.1 & 4.4
    \\
    PD $\times 100$ (a.u.)
    & 1.3 & 0.8 & 1.0 
    & 4.2 & 4.2 & 0.5
    &  24 &  24 & 0.7
    & 3.2 & 2.2 & 2.4
    & 5.6 & 5.0 & 2.5 
    \\ \hline
    & \multicolumn{15}{c|}{{\bf Test 1.2 (brain phantom)}}
    \\  \hline                                           
    T1 (\%)
    & 3.2 & 0.6 & 3.1
    & 4.5 & 2.3 & 3.3
    &  14 &  13 & 4.5
    &  11 & 1.6 &  10
    &  30 & 7.4 &  25
    \\
    T2 (\%)
    & 6.0 & 1.1 & 5.9
    & 7.7 & 4.6 & 5.1
    &  22 &  20 & 4.4
    &  22 & 5.2 &  21
    &  28 & 7.6 &  25
    \\
    PD $\times 100$ (a.u.)
    & 4.1 & 0.8 & 4.1
    & 5.9 & 3.5 & 3.8
    &  19 &  18 & 4.5
    & 9.8 & 2.3 & 9.3
    &  22 & 5.8 &  20
    \\ \hline   
    \multicolumn{16}{l}{The values in the table indicate the RMS of the total error and the partial contributions
    $\epsilon_1$ and $\epsilon_2$.}\\ 
    \multicolumn{16}{l}{For readibility, the RMSE of the proton density is multiplied by 100.}   
  \end{tabular}
  
\end{table*}

\section{In-depth analysis}
\label{sec:in-depth_analysis}

To better understand how the choice of acquisition parameters affects the reconstructions, we will perform a more detailed
analysis of the error terms $\epsilon_1$ and $\epsilon_2$ as modeled by
Eqs. (\ref{eq:theta1_with_errors}) and
(\ref{eq:theta1_with_errors_varrho0}).  
We will take a Fourier domain perspective to directly connect the predicted errors with the data acquisition process. Note that, in the Fourier domain, the convolutions present in equations (\ref{eq:define_errors_E1_E2_E3}) and (\ref{eq:define_errorsE_varrho0})
become multiplications.

\subsection{The contrast independent error term  $\epsilon_1$ and the role of variable density sampling}
\label{subsec:in-depth_contrast_independent}
We are going to show that the contrast independent error
term $\epsilon_1$ is closely related to the sampling density around the center of the $k$-space.
This error term is best modeled using variable
$\rho_0$ as in (\ref{eq:theta1_with_errors_varrho0}) and is then
given by
$|\rho_0^*(x)|^{-2} ( \Re N )^{-1} E_1(x)$ with $E_1$ as defined in 
(\ref{eq:define_errorsE_varrho0}).
 In the Fourier domain, the term
$S_{{\rm resid};p}^{(1,0)} \ast \rho_0(x)$ becomes a multiplication
between $\widehat{S}_{\rm resid}^{(1,0)}(k)$ and
$\widehat{\rho}_0(k)$. First of all, note that
Eqs.\ (\ref{eq:define_S0_11_and_S1_11}) and
(\ref{eq:resid_index_Fourier})
imply that frequent sampling at certain values
of $k$ leads to smaller values of the coefficients
$\widehat{S}_{\rm resid}^{(1,0)}(k)$ at such $k$, 
cf.\ section~\ref{subsec:sequence_properties_dependence}. Furthermore, 
$| \widehat{\rho}_0(k) |$ typically attains its largest values around $k=0$. Therefore, to minimize $\epsilon_1$
the  weights
$\widehat{S}_{\rm resid}^{(1,0)}(k)$ should be small at $k \approx 0$, which is equivalent to frequent sampling at the center of $k$-space. Since $|\widehat{\rho}_0(k) |$ decays as $\sim 1/|k|$ for large $|k|$ (a standard result from
convergence of Fourier series for piecewise continuous functions), larger values of $\widehat{S}_{\rm resid}^{(1,0)}(k)$ for large $k$ are allowed, leading to sparser sampling in the outer $k$-space region.
In conclusion, employing a scheme which frequently samples the center
of $k$-space leads to a small contrast-independent error
contribution. With radial and spiral sampling, this naturally
occurs. On the other hand, in regular Cartesian undersampling the
point $k=0$ is sampled once every $m_2 / N_{\rm US}$ times just like
other values of $k$; in this case, larger values for the contrast
independent error can be expected. This explains the large errors for
Cartesian acquisition observed in section~\ref{sec:tests}.

\subsection{The contrast dependent error $\epsilon_2(x)$ and parameters cross-talk}
\label{subsec:in-depth_contrast_dependent}

The error term $\epsilon_2(x)$ depends linearly on the contrast
$\theta_1(x)$, being the linear term in a Taylor expansion.
This means that errors in the reconstructed $T_1$ and $T_2$  depend on the true $T_1$
  and $T_2$ value maps and can thus inherit size and structure from
  them. In particular, one can expect cross-talk effects, which take place when the true value of $T_1$  influences the reconstruction of $T_2$ and vice versa. Here we will study in detail this phenomenon.
To this aim, we consider the errors $\epsilon_2$, according to Eq.\ (\ref{eq:theta1_with_errors}), which are given by $\epsilon_2(x) = (\Re N )^{-1} E_2(x)$
with $E_2(x)$ as given in Eq.\ (\ref{eq:define_errors_E1_E2_E3}).
In the Fourier domain, there is thus a simple linear relation between the
errors $\widehat{\epsilon}_2$ and the true contrast $\widehat{\theta}_1$:
\begin{equation}
  \widehat{\epsilon}_2(k) =
  \mathcal{E}_2(k) \widehat{\theta}_1(k) ,
\end{equation}
where the $N_{\rm P} \times N_{\rm P}$ matrix
$\mathcal{E}_2(k)$ is given by
\begin{equation} \label{eq:define_calE}
  \mathcal{E}_2(k)_{p,q} =
  \sum_r (\Re N)_{p,r}^{-1} \, \widehat{S}_{\text{resid};r,q}^{(1,1)}(k) .
\end{equation}

The off-diagonal coefficients of $\mathcal{E}_2(k)$ are direct
indicators of cross-talk errors and
only depend on the RF pulse sequence and $k$-space sampling scheme, 
 not on the object being scanned.
Therefore we will display some values of the $2 \times 2$ matrix block
corresponding to the parameters $\log T_1$ and $\log T_2$.
We will initially consider radial golden-angle $k$-space sampling.

As a first illustration, consider the checkerboard phantom and the
images in Fig.~\ref{fig:errormodel_checkerboard_1}. This
phantom is characterized by large $k$-space components (not shown) at $k = (\pm 0.08\pi, \pm 0.08\pi)$.  For
these values of $k$, the matrix $\mathcal{E}_2(k)$ is reported
in Table~\ref{tab:some_error_data}, at different undersampling
rates. First of all, note that as the undersampling factor increases,
also the entries of $\mathcal{E}_2(k)$ increase (in absolute
value). This is supported by the basic intuition that the larger the
undersampling, the larger the artifacts will be. Furthermore, the
large (2,1) component of $\mathcal{E}_2(k)$ for $N_{\rm US} = 32$ show that the $T_2$ reconstruction
is likely to receive a strong imprint from the true $T_1$ contrast. Indeed, in Figure~\ref{fig:errormodel_checkerboard_1} (second row, error columns) the
overestimation of the $T_2$ contrast is clearly visible in the form of a 2D sinusoidal pattern.

\begin{figure}
  \begin{center}
    \includegraphics[width=88mm]{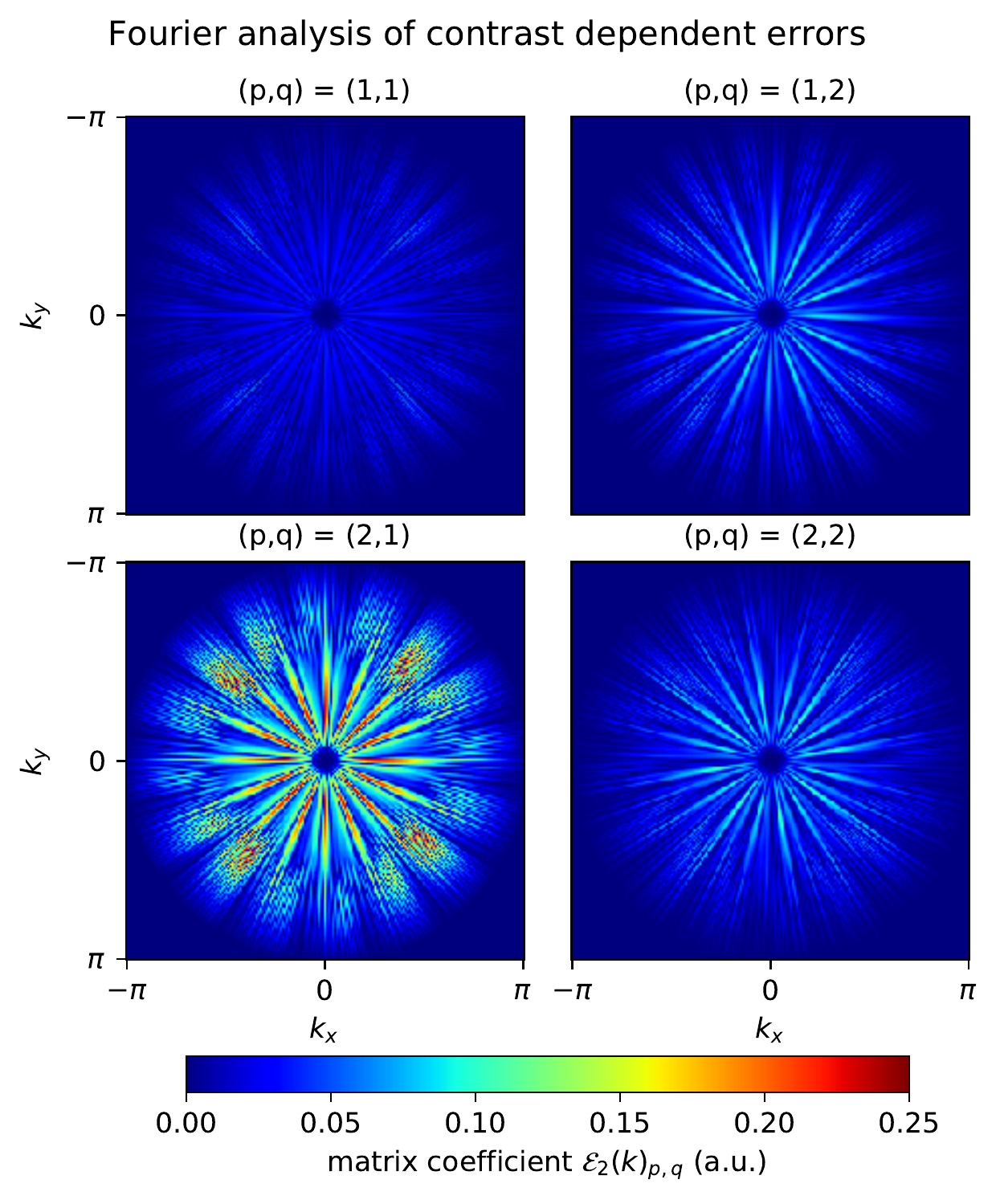}
  \end{center}
  \caption{Fourier analysis of contrast dependent errors: Plots of $\mathcal{E}_2(k)_{p,q}$ as a function of $k$ for $p,q = 1,2$. Note the relatively large coefficients for $(p,q)=(2,1)$. These lead to parameter cross-talk, mapping $T_1$ contrast in the true object to $T_2$ contrast in the reconstructions.}
  \label{fig:newFig1}
\end{figure}

In general, $\mathcal{E}_2(k)$ strongly depends on $k$, and its values for specific $k$ provide only limited information. Therefore we included plots of the absolute value of $\mathcal{E}_2(k)_{p,q}$ as a function of $k$ for $N_{\rm US} = 32$, see Figure~\ref{fig:newFig1}.
In addition, the second line of Table II contains RMS values of the errors $\mathcal{E}_2(k)_{p,q}$ for different values of $N_{\rm US}$. 
All these data show that the off-diagonal $(2,1)$ components are relatively large. This indicates that the reconstructed $T_2$ maps will be strongly affected by the actual $T_1$ maps.

In conclusion, we have already noted (see Table \ref{tab:rmsErr}) that RMS values are in general
larger for $T_2$ maps. In this section we have shown that this is partially due to cross-talk effects, which can be
severe especially for the transverse relaxation rate.

\begin{table*}
  \footnotesize\centering
  \caption{cross-talk error matrix $\mathcal{E}_2(k)$}
  \label{tab:some_error_data}
  \begin{tabular}{|l|cccc|} \hline 
    & $N_{\rm US} = $ 8 & 16 & 32 & 64
    \\ \hline
    \parbox{42mm}{Example: $k = (0.08,0.08) \pi$, errors}
    &
             $\begin{pmatrix}
               0.00  & 0.01 \\
               0.03  & 0.01
             \end{pmatrix}$ &
             $\begin{pmatrix}
               -0.03 & 0.06 \\
               0.17  & -0.04
             \end{pmatrix}$ &
             $\begin{pmatrix}
               -0.03 & 0.05 \\
               0.17  & -0.04
             \end{pmatrix}$ &
             $\begin{pmatrix}
               -0.01 & 0.07 \\
               0.18  & -0.02
             \end{pmatrix}$
    \\ \hline
    \parbox{42mm}{Golden angle k-space sampling\\ $|k|$=0.25$\pi$, RMS
    errors} &
                              $\begin{pmatrix}
                                0.00  & 0.03 \\
                                0.05  & 0.03
                              \end{pmatrix}$ &
                                              $\begin{pmatrix}
                                                0.02  & 0.05 \\
                                                0.12  & 0.07
                                              \end{pmatrix}$ &
                                                              $\begin{pmatrix}
                                                                0.02  & 0.05 \\
                                                                0.12  & 0.07
                                                              \end{pmatrix}$ &
                                                                              $\begin{pmatrix}
                                                                                0.07  & 0.07 \\
                                                                                0.13  & 0.17
                                                                              \end{pmatrix}$
    \\ \hline
    \parbox{42mm}{Random k-space sampling\\ $|k|$=0.25$\pi$, RMS errors}
    & 
             $\begin{pmatrix}
               0.13  & 0.05 \\
               0.22  & 0.16
             \end{pmatrix}$ &
             $\begin{pmatrix}
               0.21  & 0.07 \\
               0.35  & 0.22
             \end{pmatrix}$ &
             $\begin{pmatrix}
               0.30  & 0.10 \\
               0.56  & 0.32
             \end{pmatrix}$ &
             $\begin{pmatrix}
               0.43  & 0.15 \\
               0.76  & 0.46
             \end{pmatrix}$
    \\ \hline    
  \end{tabular}
\end{table*}

\subsection{The role of randomness  and
  the type of RF excitation.}
\label{subsec:sequence_properties_dependence}

Since its conception, randomness has been
 a fundamental component of the MRF framework. The general
understanding is that randomness in $k$-space sampling and/or RF
excitation trains promotes richness of encoding and better
reconstructions.  To test this assumption, a similar analysis as in
the previous subsection is performed for a radial $k$-space sampling
scheme in which the angles are randomly permuted.  RMS averaged
values of $\mathcal{E}_2(k)$ (over $k$ values and random realizations)
are given in the third row of Table~\ref{tab:some_error_data} and are
much larger than the previous, golden angle scheme. We
therefore expect the
performance of MRF reconstruction to be considerably worse for this
randomized acquisition scenario. Our prediction is confirmed by the RMS error values obtained from
this scheme, which are included in Table~\ref{tab:rmsErr}. Clearly,
random $k$-space sampling does  not necessarily lead to good MRF
imaging.

To understand the reason for this, we investigate the dependence
of the $\widehat{S}^{(\alpha,\beta)}(k)$ on the sampling scheme.
In analogy to Eq. (\ref{eq:define_calE}), let's consider the factor $\sum_r (\Re N)^{-1}_{p,r} \, \widehat{S}_{r,q}^{(1,1)}(k)$ for some fixed $k$, and write
\begin{equation} \label{eq:define_f_g}
  \sum_r (\Re N)^{-1}_{p,r}  \, \widehat{S}_{r,q}^{(1,1)}(k)
  = \langle f ,g_{p,q} \rangle
\end{equation}
where, to simplify the notation, we define
$f,g \in \CC^{N_{\rm I}}$ having components
$f(j) = \widehat{P}_j(k)$ and 
$g_{p,q}(j) = \sum_r (\Re N)^{-1}_{p,r} \mathcal{D}M(\theta_0)_{j;r} 
\overline{ \mathcal{D} M(\theta_0)_{j;q} }$.
In other words, the error terms are decomposed into a $k$-space
sampling dependent part ($f$) and an RF pulse train dependent part
($g_{p,q}$).
Let $\widetilde{f}(\nu)$ denote the Fourier transform
of $f$, given by
$\widetilde{f}(\nu)
= \sum_{j=1}^{N_{\rm I}} f(j) e^{-2\pi i \nu (j-1)}$,
$\nu = 0,\ldots, N_{\rm I}-1$, and similar
for $\widetilde{g}_{p,q}$.
From elementary Fourier theory it follows that
\begin{equation}
  \sum_r (\Re N)_{p,r}^{-1} \, \widehat{S}_{r,q}^{(1,1)}(k)
  = \frac{1}{N_{\rm I}} \langle \widetilde{f} ,\widetilde{g}_{p,q} \rangle .
  \label{eq:sum}
\end{equation}
In addition,   from Eq.\ (\ref{eq:define_P_from_Pj}) and Eq.\ 
(\ref{eq:define_S0_11_and_S1_11}) we have that
\begin{equation}
  \sum_r (\Re N)_{p,r}^{-1} \, \widehat{S}_{{\rm mean};r,q}^{(1,1)}(k)
  = \frac{1}{N_{\rm I}} \widetilde{f}(0) \overline{\widetilde{g}_{p,q}(0)} .
\end{equation}
Therefore
$\sum_r (\Re N)_{p,r}^{-1} \, \widehat{S}_{{\rm resid};r,q}^{(1,1)}(k)$
is given by the sum in Eq. (\ref{eq:sum}) where $\nu = 0$ is omitted:
\begin{equation} \label{eq:resid_index_Fourier}
  \sum_r (\Re N)_{p,r}^{-1} \, \widehat{S}_{{\rm resid};r,q}^{(1,1)}(k)
  = \frac{1}{N_{\rm I}}
  \sum_{\nu = 1}^{N_{\rm I}-1}
  \widetilde{f}(\nu) \overline{\widetilde{g}_{p,q}(\nu)} .
\end{equation}
  
An inspection of these Fourier transform terms (See Supplementary
material
section~5)
reveals the following behavior for $\widetilde{f}$ and the
$\widetilde{g}_{p,q}$ in the case of radial sampling and RF pulse sequence 1. A large part of the energy of the $\widetilde{g}_{p,q}$
(the RF excitation dependent terms) is contained in the diagonal coefficients
(i.e.\ matrix indices $p = q$)  with $\nu =0$,
while the higher Fourier coefficients decay rapidly as a
consequence of the smoothness of the magnetization response
(Fig.~S2(a)).  At the same time, the
energy in $\widetilde{f}$ (the $k$-space sampling dependent term) is
concentrated in a few, regularly spaced peaks
as a consequence of the highly
structured, golden angle $k$-space sampling scheme. The distance
between these peaks is such that $\widetilde{g}$ is already
negligible at the peak locations
with $\nu \neq 0$ (Fig.~S3(a)).  Therefore, the sum
(\ref{eq:resid_index_Fourier}) and the corresponding error term are
relatively small.
Suppose now that the $k$-space sampling
scheme is replaced by its randomized version. In this case,
larger values of $\widetilde{f}$ at low but nonzero $\nu$ lead to larger
values of the sum in Eq.\ (\ref{eq:resid_index_Fourier})
(Fig.~S3(b)). This explains the
larger errors for the
random sampling.

\begin{figure}
  \begin{center}
    \includegraphics[width=7.0cm]{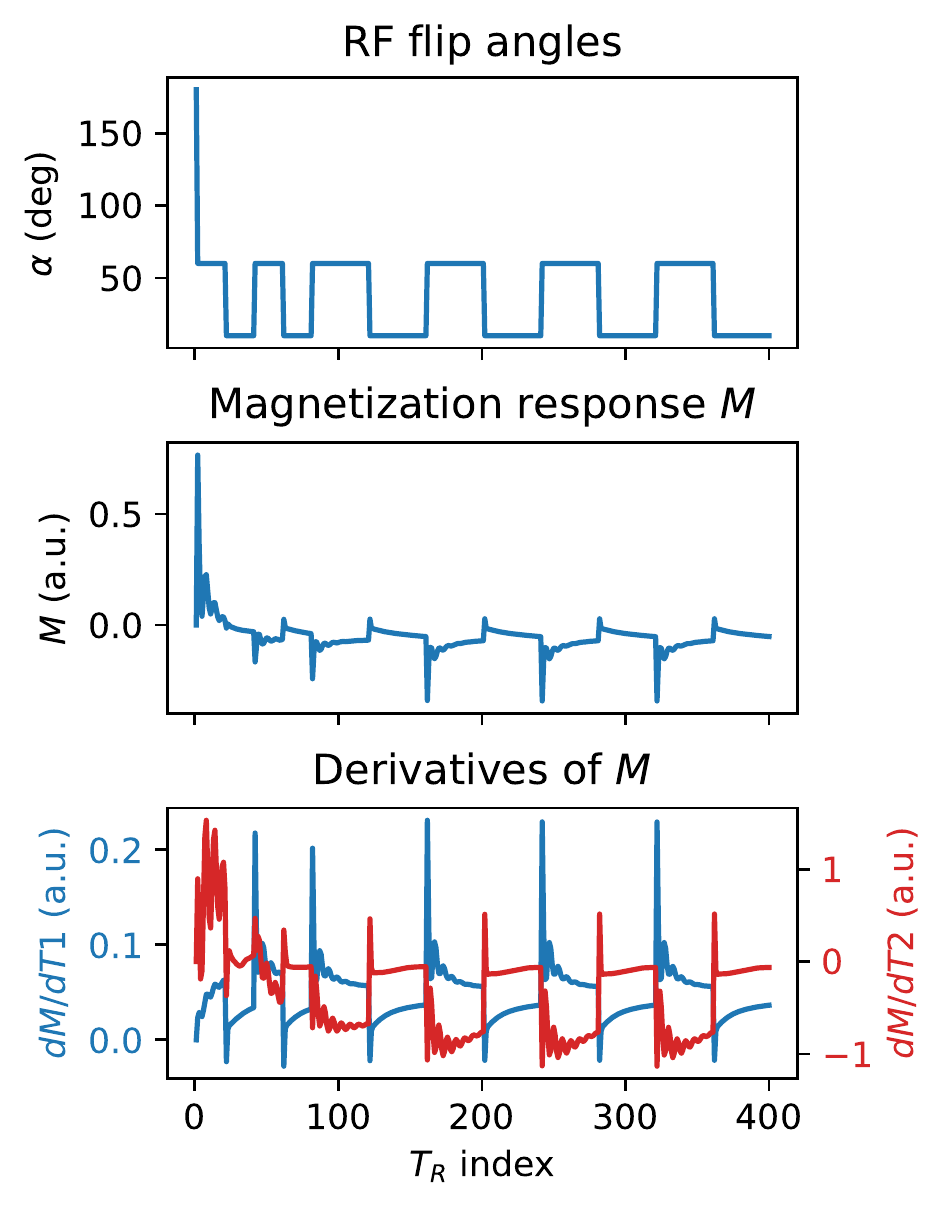}
  \end{center}
  \caption{Flip angles and plot of $M(t,\theta)$ and some derivatives
    for $T_1 = 1.0$ s, $T_2 = 0.08$ s, PD = 1 (a.u.), for sequence 2.}
  \label{fig:SeqInfo_Seq2}
\end{figure}

Let us now consider the RF dependent term, $g$, for a different flip angle train. To this aim, we introduce a new sequence, called sequence 2, which is displayed in Fig.~\ref{fig:SeqInfo_Seq2}.
The RMS error values obtained for this sequence are given in the right
section of Table~\ref{tab:rmsErr}.
According to these results, the non-smooth Sequence 2 leads to larger MRF reconstruction errors in the tissue parameters. This fact is easily explained at the hand of
Eq.\ (\ref{eq:resid_index_Fourier}). Non-smooth magnetization responses
have slower decay of Fourier coefficients. This will naturally lead to
larger values of Eq.\
(\ref{eq:resid_index_Fourier}), which is exactly what we observe for
sequence 2 (Fig.~S2(b)).

In conclusion, Eq.\ (\ref{eq:resid_index_Fourier}) suggests that the RF pulse train and the $k$-space sampling scheme should destructively interfere (or, in mathematical terms, be orthogonal) in the Fourier domain. 
This fact is reminiscent of the incoherence
between encoding and sparsity transforms in the compressed sensing
framework \cite{lustig2008compressed}. Temporal randomness in either
$k$-space or RF excitation is just a particular way to achieve
incoherence in MRF. In fact, as it has been shown in recent work
\cite{asslander2017relaxation,zhao2017optimal,sbrizzi2017dictionary},
randomly perturbed sequences do not have additional value with respect
to encoding performance and smooth RF trains can indeed perform very
well. The analysis from this section provides an explanation for this
fact and a criterion for optimizing the acquisition protocol.

\section{Discussion}\label{sec:discussion}
We have provided a  mathematical analysis of the combined effects of
$k$-space sampling and RF transient state excitation in the error of
MRF reconstructions. The perturbations can be
decomposed, to a first order approximation, into two main terms
($\epsilon_1$ and $\epsilon_2$) whose structure and behavior have been analyzed. Numerical simulations for different acquisition
strategies from Section \ref{sec:tests} show that our  model is
accurate and can predict the actual reconstruction error even for
realistic anatomies (brain). An in-depth analysis performed in Section
\ref{sec:in-depth_analysis} revealed that inter-parameter cross-talk
can be a substantial issue,
especially for the $T_2$ values. Furthermore, randomness in $k$-space
and/or RF excitation train appears to play a  secondary role or to be
even sub-optimal.  As a culmination of our analysis, we illustrated
how the RF and $k$-space sampling interact and jointly contribute to
the reconstruction error. An incoherence criterion for improved MRF
protocol is outlined which ensures destructive interference of the
corresponding two terms in the Fourier domain.

In section~\ref{sec:analysis1}, a model for the MRF estimate $\theta^*$
was obtained by expressing the parameters $\theta$ as a first order expansion $  \theta(x) = \theta_0 + \theta_1(x)$
where $\theta_0$ is treated as a constant or as a binary mask.
We have shown that when $\theta_0$ is close to $\theta$, our model performs very well. For brain imaging,
$T_1$ and $T_2$ values of white and gray matter are in fact relatively
close to the average of the two which can be used as reference
value. Note that gray and white matter are critical tissues for brain imaging. On the other hand, cerebrospinal fluid (CSF)  
has relatively much larger  $T_1$ and $T_2$ values but we have shown that the proposed approach is still able to model the error in a satisfactory way.
\indent The primary reason for the derivation of our error model is  to provide insights in the working of MRF. Compared to direct voxel-by-voxel dictionary match, our  model makes it possible to analyze and reveal important MRF features in a generalized context. In particular, the use of the functions
(convolution kernels) $\widehat{S}_1^{(\alpha,\beta)}(k)$
allows to draw conclusions valid, simultaneously, for all object
parameters and all kind of sequence choices such as RF train,
$k$-space coverage and under-sampling strategies. For example, in
section~\ref{sec:in-depth_analysis}, we have formally derived the
beneficial effect of frequent sampling in the center of
$k$-space. This result might not surprise the reader since it is
somehow intuitively understood. On the other hand, we have shown why
Cartesian sampling is to be avoided in MRF, a fact which explains the
few applications of Cartesian schemes in this paradigm.

Probably more interesting are the results obtained regarding the role
of randomness and irregularity in the sequence design. We have shown
that a well designed sequence does not necessarily need to include
randomization and in fact this feature could degrade the performance
of the method;  the RF train envelope and $k$-space sampling scheme
should be mutually destructively interfering in the Fourier
domain. One straightforward way to achieve this is to employ a regular
golden angle radial (or spiral) trajectory with a smooth RF excitation
train. However, this is only a    possibility and more efficient
combinations could be found. Further investigation into this direction
would go beyond the scope of this paper and is left to future studies.

Our error model, in combination with a signal term and possibly other
hardware constraints,  can be leveraged also for algorithmic
optimization of the  sequence; parameters such as the number of
acquisition intervals (snapshots), the values of the RF flip angles, the type of $k$-space
sampling scheme, the echo-time and repetition time, whether to use
gradient spoiling or not,  jointly and directly influence our error
model and thus can be effectively optimized at once. We believe that
this is what distinguishes our approach to previous sequence design
work where either the $k$-space sampling is not taken into account or
it is handled separately from the RF pulse design. In addition, we stress the fact that a pre-computed dictionary is not needed since the dependency of the error on the sequence parameters can be quantified purely at the hand of our model. This is a fundamental advantage for iterative sequence optimization which otherwise would require the construction of a dictionary for each new choice of sequence parameters.

As the method is based on first order Taylor expansion, questions regarding the effects of higher order terms might arise. However, while it is possible to include some higher order terms, it is not clear that this will lead to substantially better error estimates, since the convergence of the perturbation expansion is not guaranteed.

In this study we have focused on the mathematical analysis of
the MRF framework. Since a ground truth is required for error
quantification, the validation and interpretation of our model were
carried out at the hand of numerical simulations on realistic models
and scenarios. We believe the results from the numerical tests provide
sufficient explanation and  illustration of the theoretical
findings. Therefore, acquired in-vivo  data from MRI systems was not
taken into consideration.

A reader familiar with the field of inversion theory might expect such an approach to the analysis of  MRF reconstructions. Hoewever, although there are some developments towards the application of inverse theory in multi-parametric quantitative MRI (see \cite{DaviesEtAl2014_CS_MRF,asslander2018low,zhao2017optimal,sbrizzi2017dictionary,sbrizzi2018fast}), the dictionary-match approach is still the most adopted. In other words, we are interested in the mainstream implementation of MRF, which, from a mathematical perspective, is probably more challenging than the inverse problem theory. While extensive theoretical results are available for inversion problems in general (and with this we include the parametric reconstruction for Gaussian distributed noise), there is very limited work which addresses the sensitivity of the dictionary match to the case of non-Gaussian artifacts. With this work, we aim at filling this gap. An analysis of MRF for an inversion approach would result into a rather different methodology and thus it would go beyond the scope of this paper.

The popularity of MRF is mainly a consequence of its good empirical performance. We hope that this work will inspire researchers in the field to apply our analysis to other scenarios. In particular, several extensions to this work can be investigated which could not find place in our study. For instance, the signal model can be modified to include diffusion effects, transmit RF system inhomogeneity, slice profile response \cite{ma2017slice}, balanced gradient trajectories. A thorough understanding of MRF from a theoretical point of view is necessary to pave the way for its application in the clinical setting.  This work could represent a step in this direction.
\bibliographystyle{unsrt}
\bibliography{mrirefs}

\begin{thebibliography}{10}

\bibitem{MaEtAl2013_MagneticResonanceFingerprinting}
Dan Ma, Vikas Gulani, Nicole Seiberlich, Kecheng Liu, Jeffrey~L Sunshine,
  Jeffrey~L Duerk, and Mark~A Griswold.
\newblock Magnetic resonance fingerprinting.
\newblock {\em Nature}, 495(7440):187--192, 2013.

\bibitem{JiangEtAl2015_MRFingerprinting}
Yun Jiang, Dan Ma, Nicole Seiberlich, Vikas Gulani, and Mark~A Griswold.
\newblock {MR} fingerprinting using fast imaging with steady state precession
  ({FISP}) with spiral readout.
\newblock {\em Magnetic resonance in medicine}, 74(6):1621--1631, 2015.

\bibitem{cloos2016multiparametric}
Martijn~A Cloos, Florian Knoll, Tiejun Zhao, Kai~T Block, Mary Bruno, Graham~C
  Wiggins, and Daniel~K Sodickson.
\newblock Multiparametric imaging with heterogeneous radiofrequency fields.
\newblock {\em Nature Communications}, 7:12445, 2016.

\bibitem{DaviesEtAl2014_CS_MRF}
Mike Davies, Gilles Puy, Pierre Vandergheynst, and Yves Wiaux.
\newblock A compressed sensing framework for magnetic resonance fingerprinting.
\newblock {\em SIAM Journal on Imaging Sciences}, 7(4):2623--2656, 2014.

\bibitem{McGivneyEtAl2014_SVDCompression_MRFingerprinting}
Debra~F McGivney, Eric Pierre, Dan Ma, Yun Jiang, Haris Saybasili, Vikas
  Gulani, and Mark~A Griswold.
\newblock {SVD} compression for magnetic resonance fingerprinting in the time
  domain.
\newblock {\em IEEE Transactions on Medical Imaging}, 33(12):2311--2322, 2014.

\bibitem{doneva2017matrix}
Mariya Doneva, Thomas Amthor, Peter Koken, Karsten Sommer, and Peter
  B{\"o}rnert.
\newblock Matrix completion-based reconstruction for undersampled magnetic
  resonance fingerprinting data.
\newblock {\em Magnetic resonance imaging}, 41:41--52, 2017.

\bibitem{asslander2018low}
Jakob Assl{\"a}nder, Martijn~A Cloos, Florian Knoll, Daniel~K Sodickson,
  J{\"u}rgen Hennig, and Riccardo Lattanzi.
\newblock Low rank alternating direction method of multipliers reconstruction
  for {MR} fingerprinting.
\newblock {\em Magnetic Resonance in Medicine}, 79(1):83--96, 2018.

\bibitem{zhao2016maximum}
Bo~Zhao, Kawin Setsompop, Huihui Ye, Stephen~F Cauley, and Lawrence~L Wald.
\newblock Maximum likelihood reconstruction for magnetic resonance
  fingerprinting.
\newblock {\em IEEE Transactions on Medical Imaging}, 35(8):1812--1823, 2016.

\bibitem{sbrizzi2017dictionary}
Alessandro Sbrizzi, Tom Bruijnen, Oscar van~der Heide, Peter Luijten, and
  Cornelis~AT van~den Berg.
\newblock Dictionary-free {MR} {F}ingerprinting reconstruction of
  balanced-{GRE} sequences.
\newblock {\em arXiv preprint arXiv:1711.08905}, 2017.

\bibitem{greengard2004accelerating}
Leslie Greengard and June-Yub Lee.
\newblock Accelerating the nonuniform fast {F}ourier transform.
\newblock {\em SIAM review}, 46(3):443--454, 2004.

\bibitem{fessler2003nonuniform}
Jeffrey~A. Fessler and Bradley~P. Sutton.
\newblock Nonuniform fast {F}ourier transforms using min-max interpolation.
\newblock {\em IEEE Transactions on Signal Processing}, 51(2):560--574, 2003.

\bibitem{cauley2015fast}
Stephen~F Cauley, Kawin Setsompop, Dan Ma, Yun Jiang, Huihui Ye, Elfar
  Adalsteinsson, Mark~A Griswold, and Lawrence~L Wald.
\newblock Fast group matching for {MR} fingerprinting reconstruction.
\newblock {\em Magnetic Resonance in Medicine}, 74(2):523--528, 2015.

\bibitem{chen2016mr}
Yong Chen, Yun Jiang, Shivani Pahwa, Dan Ma, Lan Lu, Michael~D Twieg,
  Katherine~L Wright, Nicole Seiberlich, Mark~A Griswold, and Vikas Gulani.
\newblock {MR} fingerprinting for rapid quantitative abdominal imaging.
\newblock {\em Radiology}, 279(1):278--286, 2016.

\bibitem{hamilton2017mr}
Jesse~I Hamilton, Yun Jiang, Yong Chen, Dan Ma, Wei-Ching Lo, Mark Griswold,
  and Nicole Seiberlich.
\newblock {MR} fingerprinting for rapid quantification of myocardial {T1},
  {T2}, and proton spin density.
\newblock {\em Magnetic Resonance in Medicine}, 77(4):1446--1458, 2017.

\bibitem{zhao2017optimal}
Bo~Zhao, Justin~P Haldar, Congyu Liao, Dan Ma, Mark~A Griswold, Kawin
  Setsompop, and Lawrence~L Wald.
\newblock Optimal experiment design for magnetic resonance fingerprinting:
  Cramer-rao bound meets spin dynamics.
\newblock {\em arXiv preprint arXiv:1710.08062}, 2017.

\bibitem{asslander2017relaxation}
Jakob Assl{\"a}nder, Riccardo Lattanzi, Daniel~K Sodickson, and Martijn~A
  Cloos.
\newblock Relaxation in spherical coordinates: Analysis and optimization of
  pseudo-{SSFP} based {MR}-{F}ingerprinting.
\newblock {\em arXiv preprint arXiv:1703.00481}, 2017.

\bibitem{bezanson2017julia}
Jeff Bezanson, Alan Edelman, Stefan Karpinski, and Viral~B Shah.
\newblock Julia: {A} fresh approach to numerical computing.
\newblock {\em SIAM Review}, 59(1):65--98, 2017.

\bibitem{kwan1999mri}
RK-S Kwan, Alan~C Evans, and G~Bruce Pike.
\newblock {MRI} simulation-based evaluation of image-processing and
  classification methods.
\newblock {\em IEEE Transactions on Medical Imaging}, 18(11):1085--1097, 1999.

\bibitem{lustig2008compressed}
Michael Lustig, David.~Ll Donoho, Juan~M. Santos, and John~M. Pauly.
\newblock Compressed sensing {MRI}.
\newblock {\em IEEE Signal Processing Magazine}, 25(2):72--82, 2008.

\bibitem{sbrizzi2018fast}
Alessandro Sbrizzi, Oscar van~der Heide, Martijn Cloos, Annette van~der Toorn,
  Hans Hoogduin, Peter~R. Luijten, and Cornelis~A.T. van~den Berg.
\newblock Fast quantitative {MRI} as a nonlinear tomography problem.
\newblock {\em Magnetic Resonance Imaging}, 46:56--63, 2018.

\bibitem{ma2017slice}
Dan Ma, Simone Coppo, Yong Chen, Debra~F McGivney, Yun Jiang, Shivani Pahwa,
  Vikas Gulani, and Mark~A Griswold.
\newblock Slice profile and {B1} corrections in {2D} magnetic resonance
  fingerprinting.
\newblock {\em Magnetic Resonance in Medicine}, 78(5):1781--1789, 2017.

\end{thebibliography}

\newpage%
\pagestyle{fancy}%
\fancyhf{}%
\fancyhead[C]{\Large APPENDIX: SUPPLEMENTARY MATERIAL}%


\section*{Supplementary material}

# Understanding the combined effect of *k*-space undersampling and transient states excitation in MR Fingerprinting reconstructions. *Supplementary Material*

Christiaan C. Stolk and Alessandro Sbrizzi

## 1 Perturbation theory with variable reference proton density

We proceed with the derivation of the equivalent of Eq. (22) from the main text in the case of spatially-dependent reference proton density, $\rho_0(x)$. We still assume there is a constant reference value $\eta_0$ for the other parameters. Recall from section II the notation $\theta = (\eta, \rho)$ or $\theta = (\eta, \operatorname{Re} \rho, \operatorname{Im} \rho)$.

In case of a real proton density, we express the quantitative parameters $\theta(x)$ in terms of the spatially dependent $\rho_0(x)$, the constant $\theta_0$ and spatially dependent contrast functions $\eta_1(x)$ and $\rho_1(x)$ as follows

$$\theta(x) = (\eta_0 + \eta_1(x), \quad \rho_0(x)(1 + \rho_1(x))). \tag{1}$$

Similarly, in the case of a complex proton density, we write

$$\theta(x) = (\eta_0 + \eta_1(x), \quad \operatorname{Re}(\rho_0(x)(1 + \rho_1(x))), \quad \operatorname{Im}(\rho_0(x)(1 + \rho_1(x)))). \tag{2}$$

This choice allows us to exploit the linearity of the signal with respect to the proton density. In particular, the decomposition $\rho(x) = \rho_0(x)(1 + \rho_1(x))$ makes it possible to bring the spatial dependent reference $\rho_0(x)$ outside the signal equation and treat the remaining part $1 + \rho_1(x)$ as a constant plus a spatial dependent contrast terms, which is the same scenario considered in Section III. Therefore, we wil be able to use the same definitions for $S_{p,q}^{(1,1)}(x)$ etc. from subsection III-B of the main text.

We can apply the perturbation theoretic analysis of Section III with constant reference $\theta_0$ and spatial dependent contrast term $\theta_1(x)$ defined in the following way. In case of real $\rho$ we write

$$\begin{aligned}\theta_0 &= (\eta_0, 1) \\ \theta_1(x) &= (\eta_1(x), \rho_1(x)).\end{aligned} \tag{3}$$

For complex $\rho$ we write

$$\begin{aligned}\theta_0 &= (\eta_0, 1, 0) \\ \theta_1(x) &= (\eta_1(x), \operatorname{Re} \rho_1(x), \operatorname{Im} \rho_1(x)).\end{aligned} \tag{4}$$

With these definitions, Eq. (2) from Section II is modified to

$$s_{j,l} = \sum_{x \in G_\mathrm{p}} \rho_0(x) M_j(\theta_0 + \theta_1(x)) e^{ik_{j,l} \cdot x}. \tag{5}$$

Note that only the multiplicative factor $\rho_0$ has been added to the original signal equation. This is a consequence of our choice for the definition of $\theta$ in Eq. (2) above.

In the absence of thermal noise, the undersampled images are given by

$$\tilde{I}_j(x) = P_j * (\rho_0(\cdot) M_j(\theta_0 + \theta_1(\cdot)))(x). \tag{6}$$

1

For the estimates $\theta^*(x)$ we use similar redefinitions, writing $\theta^*$ in terms of a variable reference proton density

$$\theta^*(x) = (\eta_0^* + \eta_1^*(x), \rho_0^*(x)(1 + \rho_1^*(x))). \tag{7}$$

By inserting the definition for $\theta$ and $\theta^*$ the stationarity equations become

$$\begin{aligned} 0 = \ &\mathrm{Re} \sum_{j=1}^{N_\mathrm{I}} |\rho_0^*(x)|^2 \overline{DM(\theta_0 + \theta_1^*(x))_{j;p}} M(\theta_0 + \theta_1^*(x))_j \\ &- \mathrm{Re} \sum_{j=1}^{N_\mathrm{I}} \sum_{y \in G_\mathrm{p}} P_j(x-y) \overline{\rho_0^*(x)} \, \overline{DM(\theta_0 + \theta_1^*(x))_{j;p}} \rho_0(y) M(\theta_0 + \theta_1(y))_j. \end{aligned} \tag{8}$$

In the above equation, the factor $\overline{\rho_0^*}$ can be taken out of the sum. The resulting equation can be expanded in a similar way as Eq. (15) of the main text. The formal first order expansion is

$$\begin{aligned} 0 = \mathrm{Re}\, \Bigg( \overline{\rho_0^*} \bigg[ &\rho_0^* S_{0;p}^{(1,0)} * 1(x) + \sum_q \rho_0^* \theta_{1,q}^* S_{0;p,q}^{(1,1)} * 1(x) + \sum_q \rho_0^* \theta_{1,q}^* S_{0;p,q}^{(2,0)} * 1(x) \\ &- S_{0;p}^{(1,0)} * \rho_0(x) - \sum_q S_{0;p,q}^{(1,1)} * (\rho_0 \theta_1)(x) - \sum_q \theta_{1,q}^* S_{0;p,q}^{(2,0)} * \rho_0(x) \\ &- S_{1;p}^{(1,0)} * \rho_0(x) - \sum_q S_{1;p,q}^{(1,1)} * (\rho_0 \theta_1)(x) - \sum_q \theta_{1,q}^* S_{1;p,q}^{(2,0)} * \rho_0(x) \bigg] \Bigg). \end{aligned} \tag{9}$$

In this equation, we can still choose the function $\rho_0^*(x)$. In order to obtain some cancellations, we set

$$\rho_0^*(x) = P * \rho_0(x). \tag{10}$$

Looking carefully at the definitions of $S_{0;p}^{(1,0)}$ and of $S_{0;p,q}^{(2,0)}$ it follows that the first and fourth terms in (9) cancel each other. The same holds for the third and sixth term. The remaining equation is

$$\begin{aligned} 0 = \mathrm{Re}\, \Bigg( \overline{\rho_0^*(x)} \bigg[ &\sum_q \rho_0^* \theta_{1,q}^* S_{0;p,q}^{(1,1)} * 1(x) - \sum_q S_{0;p,q}^{(1,1)} * (\rho_0 \theta_1)(x) \\ &- S_{1;p}^{(1,0)} * \rho_0(x) - \sum_q S_{1;p,q}^{(1,1)} * (\rho_0 \theta_1)(x) - \sum_q \theta_{1,q}^* S_{1;p,q}^{(2,0)} * \rho_0(x) \bigg] \Bigg). \end{aligned} \tag{11}$$

This is the perturbation theoretic expansion of the stationarity equations in case of a variable reference density $\rho_0(x)$. Equation (27) in the main text follows from Eq. (11) in an analogous way as Eq. (25) follows from Eq. (22).

## 2 Simulation details

Signal simulations are based on the Bloch equations in rotating coordinates. RF pulses are modeled using an instantaneous rotation with a given flip angle and phase, while between RF pulses and readout times the usual exponential decays are applied.

To model dephasing due to gradient spoiling, each voxel is assumed to consist of $N_\mathrm{GS}$ isochromats. Each isochromat is rotated by a different phase factor. The phase factors were distributed uniformly over the interval $[0, 2\pi]$. Here $N_\mathrm{GS} = 128$ was chosen.

Non-uniform Fourier transform reconstructions are performed by the NUFFT package of reference [1]. The fundamental issue in the image reconstruction is the choice of the $k$-space weights $w_{j,l}$. These should be set such that the resulting point spread function is a bandlimited Dirac delta function that is smoothly

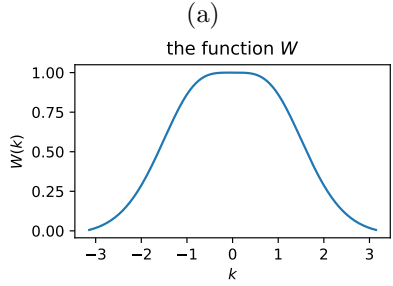

Figure 1: The smooth, low pass $k$-space windowing function employed for the reconstruction of the snapshots images.

truncated in the Fourier domain, as a hard truncation leads to oscillatory nonlocal contributions (Gibbs phenomenon). In radial sampling, we aim at

$$\widehat{P}(k) = W(|k|), \tag{12}$$

where $W$ is a smooth window function on $[-\pi, \pi]$, see Fig. 1. To obtain this, the density of $k$-space sampling must be taken into account. Relative to Nyquist sampling, this density, averaged over circles of constant $|k|$, is given by

$$\begin{aligned} \text{relativeDensity}(|k|) &= \text{relativeAngularDensity} \cdot \text{relativeRadialDensity} \\ &= \frac{2}{|k|} \frac{1}{N_{\text{US}}} \cdot \text{relativeRadialDensity} \ . \end{aligned} \tag{13}$$

Based on this, the weights $w_{j,l}$ were defined by

$$w_{j,l} = \begin{cases} W(|k_{j,l}|)/\text{relativeDensity}(|k_{j,l}|) & \text{if } k \neq 0 \\ C_{\text{R},0} & \text{if } k = 0 \end{cases} \tag{14}$$

where $C_{\text{R},0}$ is such that the Fourier transform of the PSF at $k=0$ (approximated by applying the PSF to the function 1) is equal to 1 at $x=0$. For spirally sampled $k$-space the weights were defined by

$$w_{j,l} = W(|k_{j,l}|)/\text{relativeDensitySpiral}(|k_{j,l}|), \tag{15}$$

resulting in a similar point spread function. For Cartesian sampling the weights were given by

$$w_{j,l} = W(|(k_{j,l})_1|)W(|(k_{j,l})_2|)N_{\text{US}} \tag{16}$$

In this case the point spread function was similarly well-localized as the radial and spiral ones.

## 3 Further examples for validation

The checkerboard example in section IV-B of the main text has a constant proton density and moderate variations in $T_2$. For the purpose of validation two more checkerboard examples with larger variations in these parameters are shown in Figure 2.

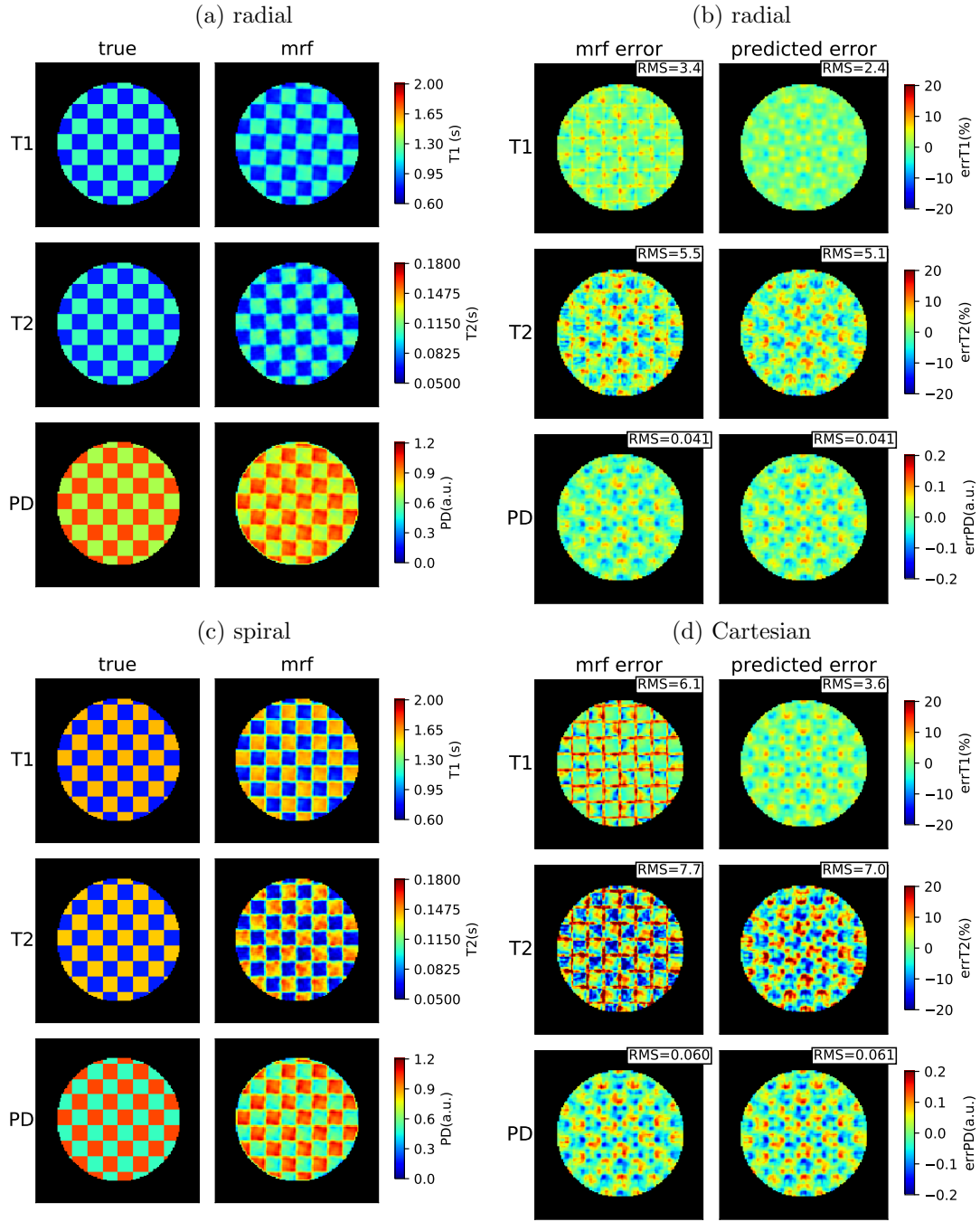

Figure 2: Numerical checkerboard phantoms with contrast factors 1.5 and 2. (a,c) true model and MRF reconstructions; (b,d) actual and predicted MRF errors for radial sampling. Parameters values in (a) $T_1$ : (0.8, 1.2) s, $T_2$ : (0.07, 0.105) s, PD : (0.67,1.0) a.u. Parameter values in (c): $T_1$ : (0.8, 1.6) s, $T_2$ : (0.07, 0.14) s, PD : (0.5,1.0) a.u.

## 4 Analysis of the undersampling errors $e_{\text{US}}$

In Figure 3 some data is shown that confirm that, in general, the undersampling errors $e_{\text{US}}$ are not independent and identically distributed Gaussian noise.

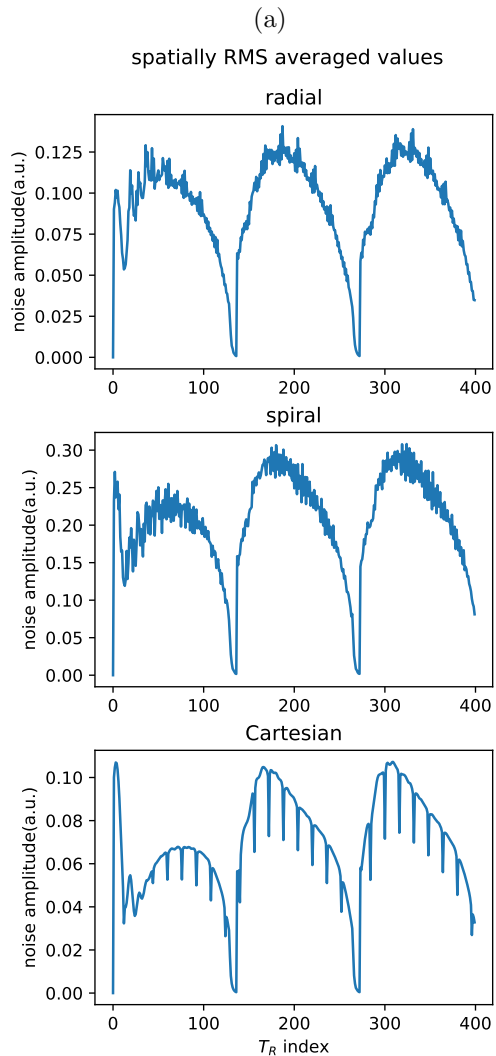

Figure 3: Spatial RMS averages of the undersampling errors $e_{\text{US},j}$ as a function of time for the sequence described in section IV-A applied to the brain phantom described in IV-C.

# 5 Figures for section V-C

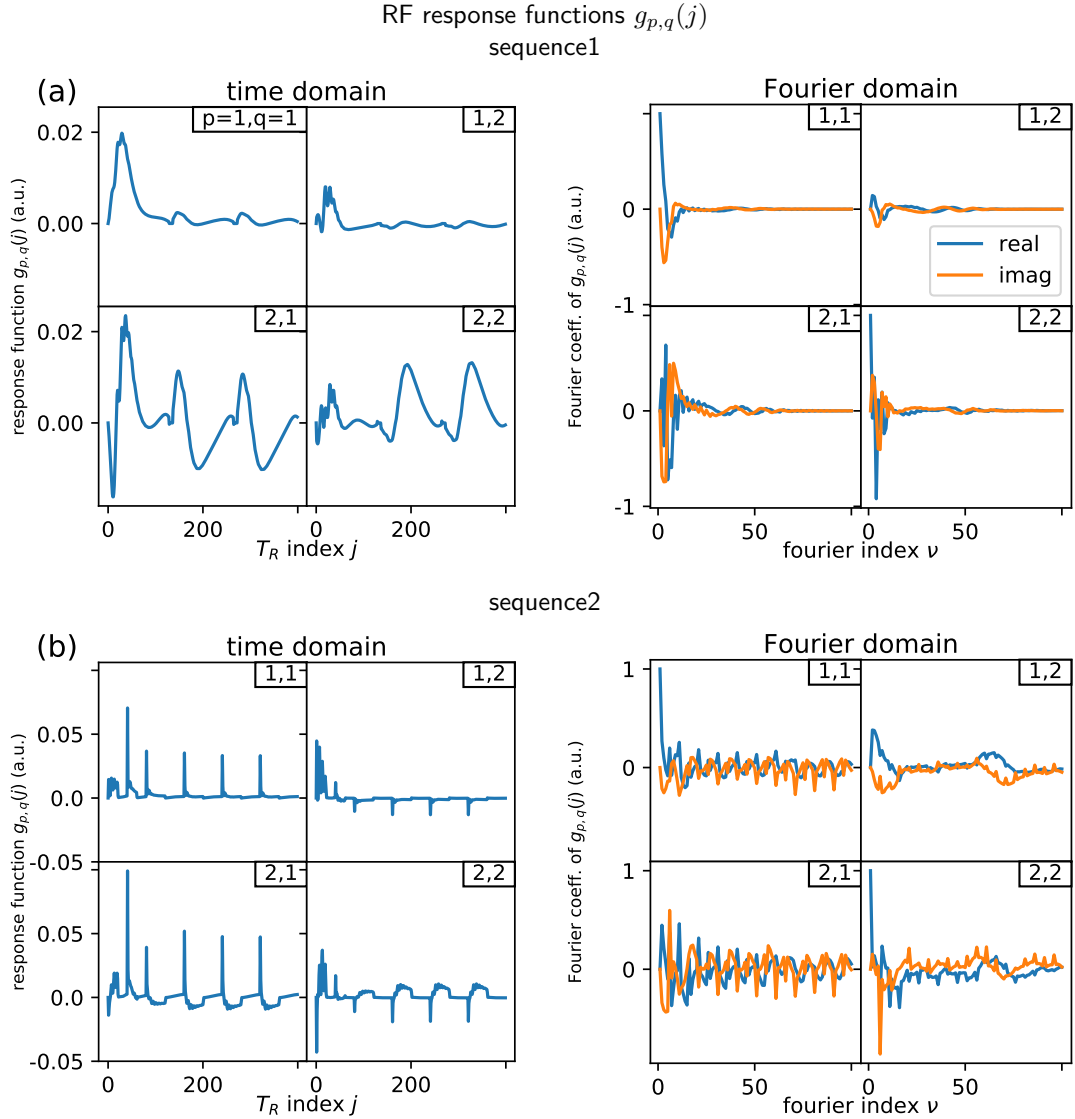

Figure 4: The time-dependent matrix components $g_{p,q}(j)$ and their Fourier transforms for (a) sequence 1 and (b) sequence 2. See section V-C. Note the rapid decay of the Fourier coefficients in sequence 1. For sequence 2, the decay is much slower and the coefficients are large also for $\nu \gg 0$. Note also that for sequence 1, the $(2,1)$ coefficient appears to be relatively large compared to the other matrix components. This fact can explain the substantial crosstalk in the reconstructed $T_2$ (see also section V-B).

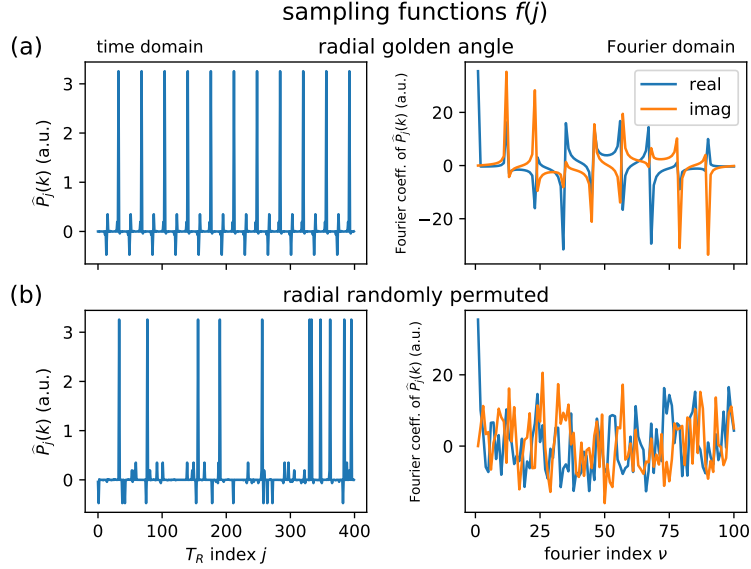

Figure 5: Plots of the time-dependent sampling schemes $f_j = \widehat{P}_j(k)$, $j = 1, \ldots, N_\mathrm{I}$ and their Fourier transform in $j$. See section V-C. Two sampling schemes are considered: (a) golden angle radial sampling and (b) randomly permuted radial sampling. The undersampling factor is 32 and a random value of $k$ with $|k| = 0.25\pi$ is chosen. The behavior is independent of the specific choice of $k$. For the golden angle sampling, the regular pattern in the time domain leads to regularly spaced peaks in the Fourier domain. For randomly permuted radial sampling, such regularly spaced peaks are absent and the Fourier transform exhibits a typical random behavior.

# References

[1] Leslie Greengard and June-Yub Lee. Accelerating the nonuniform fast Fourier transform. SIAM review, 46(3):443–454, 2004.



\end{document}